\documentclass[review,onefignum,onetabnum]{siamonline171218}
\usepackage{amsmath}
\usepackage{amsfonts}
\usepackage{amssymb}
\usepackage{bbm}
\usepackage{parskip}
\usepackage{nameref}
\usepackage{hyperref}
\usepackage{graphicx}
\usepackage{tabulary}
\usepackage[nopatch=eqnum]{microtype}
\DisableLigatures[f]{encoding = *, family = * }
\usepackage{array}
\usepackage{changepage}
\usepackage{cite}
\usepackage{textcomp,marvosym}\headers{Sampling parameters of ODEs with constrained Langevin dynamics}{C. Chi, J. Weare, A. R. Dinner}
\usepackage{algpseudocode}
\usepackage{colortbl}
\usepackage{cleveref}
\title{Sampling parameters of ordinary differential equations\\ with constrained Langevin dynamics}
\author{Chris Chi\thanks{Department of Chemistry, University of Chicago, Chicago, Illinois, USA (\email{dinner@uchicago.edu}).}
\and Jonathan Weare\thanks{Courant Institute of Mathematical Sciences, New York University, New York, New York, USA.}
\and Aaron R.\ Dinner\footnotemark[1]}

\begin{document}

\nolinenumbers

\maketitle

\section*{Abstract}
Fitting models to data to obtain distributions of consistent parameter values is important for uncertainty quantification, model comparison, and prediction.  Standard Markov chain Monte Carlo (MCMC) approaches for fitting ordinary differential equations (ODEs) to time-series data involve proposing trial parameter sets, numerically integrating the ODEs forward in time, and accepting or rejecting the trial parameter sets.  When the model dynamics depend nonlinearly on the parameters, as is generally the case, trial parameter sets are often rejected, and MCMC approaches become prohibitively computationally costly to converge.  Here, we build on methods for numerical continuation and trajectory optimization to introduce an approach in which we use Langevin dynamics in the joint space of variables and parameters to sample models that satisfy constraints on the dynamics.  We demonstrate the method by sampling Hopf bifurcations and limit cycles of a model of a biochemical oscillator in a Bayesian framework for parameter estimation, and we attain performance that matches or exceeds the performance of leading MCMC approaches that require numerically integrating the ODEs forward in time.  We describe numerical experiments that provide insight into the speedup.  The method is general and can be used in any framework for parameter estimation and model selection.

\section{Introduction}

Fitting mathematical models to data provides insight that may be difficult to obtain directly from experiments.  One can formulate models to test competing hypotheses and compare them systematically for how readily the models fit the data while accounting for their complexity (for examples from systems biology, which is our application focus here, see \cite{Battogtokh_2002,Brown_2003,Vyshemirsky_2008,Flaherty_2008,Xu_2010,Hug_2013,Eydgahi_2013,Mello_2018,Hong_2020}).  In so doing, one learns the implications of the fits for parameters and variables that are not measured, and one can interpolate the measurements to arbitrary resolution.  These predictions (and their uncertainties) can in turn be used to guide the design of additional experiments and manipulations for control (e.g., therapeutic interventions).


Typically, the data are insufficient to constrain models fully, and the parameters have a distribution of values that are consistent with the data.  An accurate representation of this distribution is important for both quantitative and qualitative assessment of the associated models.  Because a closed form expression for the distribution is rarely available, numerical methods such as Markov chain Monte Carlo (MCMC) are often used to sample the distribution.  In MCMC, one generates random sequences of parameter sets in such a way that their distribution converges to the distribution of interest.

Two features of parameter distributions often make MCMC converge prohibitively slowly.  First, they can be multimodal, and sequences of low probability parameter sets must be sampled for MCMC to transition between the modes.  
Second, parameter distributions are generally poorly scaled \cite{Gutenkunst_2007}.  That is, the support of the distribution may be much narrower in certain directions than others.  Because the narrow (stiff) and wide (soft) directions often involve combinations of multiple parameters, they are not obvious, and the narrowest direction often limits the separations between successive parameter sets and, in turn, the amount of space explored.


Much work has been done to address these issues.  Multimodality can be treated by enhancing the sampling of the low-probability sequences of parameter sets that enable transitions \cite{Dinner_2020,Matthews_2018,Neal_2001,Murakami_2014,Pullen_2014,Swendsen_1986}.  Poorly scaled distributions can be sampled more efficiently by schemes in which proposed changes to the parameters are asymmetric \cite{Cotter_2013}, affine invariant \cite{Goodman_2010, Leimkuhler_2017}, or otherwise depend on the current position in parameter space \cite{Girolami_2011, Zhang_2011}. 
Nevertheless, even with enhanced sampling and preconditioning, generally $10^6$ to $10^9$ parameter sets must be proposed and evaluated to obtain a reasonable sample of a parameter distribution.  As a result, even when a single parameter set is a fraction of a second to test, obtaining a good sample of a distribution can be prohibitively computationally costly.

Here, we develop a complementary approach based on the observation that it can be much faster to compute the limiting behavior of a (deterministic) dynamical model---for example, its fixed points and limit cycles---than to integrate it forward in time.  
Namely, we show how constrained Hamiltonian dynamics can be used to sample fixed points consistent with data directly; the fixed points can be used to seed more computationally expensive calculations to obtain transient behaviors.  
While related methods have been applied in a variety of contexts \cite{Brubaker_2012, Xu_2024}, to the best of our knowledge, they have not been previously applied to estimating parameters for systems of ordinary differential equations (ODEs). We focus particularly on ODE systems exhibiting oscillatory dynamics \cite{Hwang_2025}, which can be especially difficult to sample because freedom in the phase of oscillations causes the parameter distributions of such models to be poorly scaled and multimodal. In Section \ref{sec:repressilator}, we motivate the approach by discussing the problem of sampling parameters for a model of the repressilator, an oscillatory gene regulatory network that we introduce. In Section \ref{sec:theory}, we provide a conceptual overview of the approach. 
In Section \ref{sec:calculations}, we return to the problem of sampling parameters of the repressilator and show that, for it, we can outperform widely used MCMC methods. We conclude by discussing how the method could be extended and combined with other strategies for accelerating parameter estimation in Section \ref{sec:conclusion}. Technical details regarding our implementation of the method are given in Section \ref{sec:methods}.




\section{Motivating example}\label{sec:repressilator}


Suppose selected measurements are made on a biochemical system.  The problem that we consider is how to fit the data (or more generally how to satisfy a set of constraints on the dynamics) to evaluate models and/or estimate parameter values.
When spatial heterogeneities are not resolved and stochasticity can be neglected, the dynamics of such a system can be modeled through a set of ODEs. 

\begin{figure}[bt]
    \centering
    \includegraphics[width=0.33\textwidth]{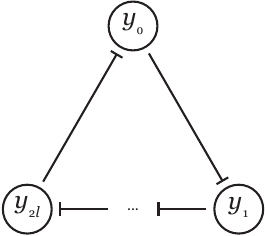}
    \caption{Schematic of the repressilator model with \(2l + 1\) species, where \(l\) is a positive integer.}
    \label{fig:repressilator_diagram}
\end{figure}

\begin{figure}[bt]
    \centering
    \includegraphics[width=0.5\linewidth]{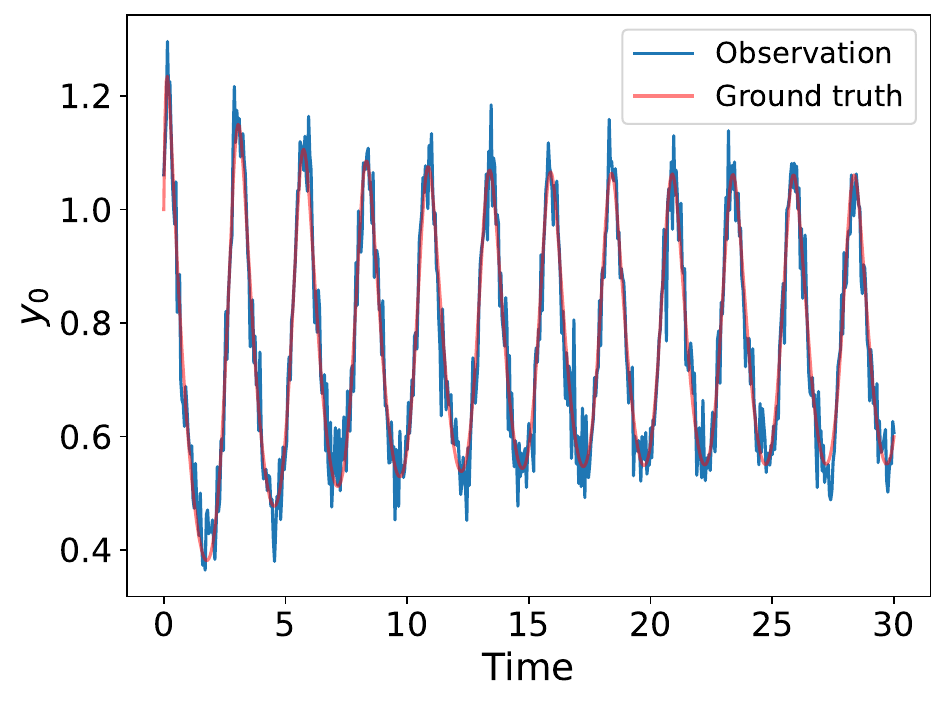}
    \caption{Representative simulated data used in parameter estimation for the three- and seven-species repressilator. For the data shown, we integrate the three-species equations with arbitrarily chosen oscillatory parameters forward in time using Tsitouras's 5/4 explicit Runge-Kutta method \cite{Tsitouras_2011} to generate the ground truth data shown in red. Simulated observations, shown in blue, are generated by adding normally distributed noise with mean zero and variance \(2.5\times10^{-3}\) to the concentration of the first species.}
    \label{fig:repressilator_data}
\end{figure}

For concreteness, consider the repressilator \cite{Elowitz-2000}, a cycle of inhibitory gene regulatory interactions that can give rise to oscillations in gene expression (Figures \ref{fig:repressilator_diagram} and \ref{fig:repressilator_data}). A minimal model of this system is the set of ODEs
\begin{subequations}\label{eqn:repressilator}
    \begin{align}
        \frac{d\mathbf y}{ds} & = \mathbf f(\mathbf y, \mathbf k, \mathbf n)\\
         f_j(\mathbf y, \mathbf k, \mathbf n)&= \frac{k_{j,0}}{1 + y_{j - 1}^{ n_{j - 1}}} - 
    \frac{k_{j,1}}{k_{0,1}}y_j,~j\in\{0, 1, \dots, 2\ell\},
    \end{align}
\end{subequations}
where \(y_j\) is the concentration of the \(j\)th gene product, \(k_{j,0}\) are dimensionless expression rates, \(k_{j,1}\) are dimensionless degradation rates, \(n_j\) are Hill coefficients, \(s\) is dimensionless time, and \(\ell > 1\) is a positive integer. The first term on the right hand side of \eqref{eqn:repressilator} represents the inhibitory effect of the previous species in the cycle on expression and the second term represents degradation of the gene product. For a fixed \(\ell\), we would like to infer values for \(\mathbf k\) and \(\mathbf n\) that are consistent with the (simulated) data in Figure \ref{fig:repressilator_data}, for example. 

Bayesian statistics provide a framework for accomplishing this goal \cite{Mackay_2003,Vyshemirsky_2008,Eydgahi_2013}: let \(\mathbf{k}\) be the parameters of the model and \(\mathbf x\) be experimental observations. Given a prior distribution \(\pi(\mathbf{k})\) and likelihood function \(\pi(\mathbf x|\mathbf{k})\), the posterior distribution \(\pi(\mathbf{k}|\mathbf x)\) can be computed using Bayes' rule:
\begin{equation}\label{eq:bayes}
\pi(\mathbf{k}|\mathbf x) = \frac{\pi(\mathbf x|\mathbf{k})\pi(\mathbf{k})}{\pi(\mathbf x)} \propto \pi(\mathbf x|\mathbf{k})\pi(\mathbf{k}).
\end{equation}
That is, the probability of the model given the data is proportional to the product of our prior belief in the model and the probability of the data given the model. Assuming that the experimental uncertainties of observable $i$ are independent and normally distributed with variance $\sigma_i^2$, the likelihood takes the form
\begin{equation}
    \pi(\mathbf x|\mathbf k) \propto \exp\left(-\sum_{i=1}^N\sum_{j=1}^{M_i}\frac{1}{2\sigma_{i}^{2}}\left(\hat x_i - x_i\right)^2\right),
\end{equation}
where \(\hat{\mathbf x}\) is the model output, $N$ is the number of observables, and $M_i$ is the number of measurements of observable $i$. As the likelihood can vary over several orders of magnitude and the logarithm is a strictly increasing function, computations are usually performed with the logarithm of the likelihood (log-likelihood). Maximizing the log-likelihood corresponds to minimizing the weighted sum of squared errors:
\begin{equation}
    \log \pi(\mathbf x|\mathbf k) = \sum_{i=1}^N\sum_{j=1}^{M_i}\frac{1}{2\sigma_{i}^{2}}\left(\hat x_i - x_i\right)^2 + C,\label{eqn:LL_lstsq}
\end{equation}
where \(C\) is a constant that is independent of \(\mathbf x\) and \(\mathbf k\).

In general, closed form expressions for the marginal likelihood \(\pi(\mathbf x)\) are not available, and numerical estimates can be difficult to compute when \(\mathbf x\) is more than a couple of dimensions. Direct computation of \(\pi(\mathbf x)\) can be avoided by using Markov chain Monte Carlo (MCMC) methods to sample sequences of parameter sets in such a way that their distribution converges to \(\pi(\mathbf{k}|\mathbf x)\). 

Random Walk Metropolis (RWM) is one of the simplest and most widely used forms of MCMC.  When applied to an ODE model like that for the repressilator above, it involves many steps of the form in Algorithm \ref{alg:mcmc}.  In words, a random change to the parameter values is made (drawn from a normal distribution in the example shown), the posterior is evaluated (which involves integrating the ODEs), and the step is accepted with probability
\begin{equation}
    \min\left\{\frac{\pi(\mathbf x|\mathbf k_{i + 1})\pi(\mathbf k_{i + 1})}{\pi(\mathbf x|\mathbf k_i)\pi(\mathbf k_i)},1\right\},
\end{equation}
where \(\mathbf k_{i}\) are the current parameter values and \(\mathbf k_{i + 1}\) are the perturbed parameter values.
\begin{algorithm}
\caption{One step of Random Walk Metropolis \label{alg:mcmc}}
\textbf{Input: }{current model parameters \(\mathbf k_i\), current posterior value \(\pi_i\), step size \(\Delta t\)}\\
\textbf{Output: }{updated model parameters \(\mathbf k_{i + 1}\), updated posterior value \(\pi_{i + 1}\)}
    \begin{algorithmic}[1]
        \State \(\eta \sim \mathcal N(0, \mathbf I)\)
        \State \(\mathbf k_{i + 1} \leftarrow \mathbf k_i + \eta\Delta t\)
        \State \(\pi_{i + 1}\leftarrow \pi(\mathbf x|\mathbf k_{i + 1})\pi(\mathbf k_{i + 1})\) \hfill\Comment{Integrate \eqref{eqn:repressilator}}\label{likelihood_calculation}
        \State \(r \sim U(0,1)\)
        \If{\(r > \pi_{i + 1} / \pi_{i}\)}
            \State \(\mathbf k_{i + 1}\leftarrow \mathbf k_{i}\)
            \State \(\pi_{i + 1}\leftarrow \pi_i\)
        \EndIf
        \State \textbf{return} \(\mathbf k_{i + 1}, \pi_{i + 1}\)
    \end{algorithmic}
\end{algorithm}

In principle, RWM provides an unbiased estimate of the target distribution when run for a sufficiently large number of steps; in practice, this number may be intractably large because the distribution may be multimodal and poorly scaled, as discussed above. 
%
There are many ways to improve the performance of MCMC, but, in the absence of a closed form expression for the likelihood, the operations in step \ref{likelihood_calculation} of Algorithm \ref{alg:mcmc} are unavoidable and usually constitute the bulk of the computational cost. Therefore, we can improve the performance of any MCMC algorithm by reducing the total amount of effort spent on step \ref{likelihood_calculation}, through reducing either the cost of evaluating the likelihood or the number of likelihood evaluations required to generate independent samples from \(\pi(\mathbf{k}|\mathbf x)\). Our method accomplishes both of these.

\section{Theory}\label{sec:theory}

The key observation that we make to improve on the efficiency of basic MCMC methods like RWM is that certain properties may be more tractable to compute when expressed as constraints. In the case of a dynamical system, limiting properties can be computed without integrating the differential equations describing the dynamics. For example, if the steady-state behavior of a system described by \(\dot{\mathbf y}(t) = \mathbf f(t)\) is of interest and the system is known to converge to a fixed point at long times, it can be several orders of magnitude cheaper to compute \(\mathbf y^*\) such that \(\mathbf f(\mathbf y^*) = \mathbf 0\) with a root-finding algorithm than to integrate \(\dot{\mathbf y}(t) = \mathbf f(t)\) for times of interest. For systems such as the repressilator with complicated limiting dynamics such an approach does not provide a complete description of the system, but it can still provide a way to characterize regions of the parameter space rapidly. 
To the extent that a system is known to satisfy certain constraints, it can be better to introduce them directly through equations of the form \(\mathbf f(\mathbf y) = \mathbf 0\) than indirectly through penalty terms added to the log-likelihood because the latter strategy tends to make the support for the target distribution narrow and limit the step size, as discussed above.
Thus our goal is to develop a way to sample parameter values that satisfy constraints of the form \(\mathbf f(\mathbf y) = \mathbf 0\).


More precisely, suppose that some properties of a system can be expressed as a system of equations \(\mathbf f(\mathbf y, \mathbf k) = \mathbf 0\), where \(\mathbf f:\mathbb R^{d}\times\mathbb R^{n}\rightarrow \mathbb R^{m}\), \(\mathbf y\in\mathbb R^d\) describes the state of the system, and \(\mathbf k\in\mathbb R^n\) is a vector of adjustable parameters. Our goal is to sample points from the solution set \(\mathcal M \equiv \{(\mathbf y,\mathbf k)\, |\, \mathbf f(\mathbf y,\mathbf k)=\mathbf 0\}\) efficiently. We show how numerical continuation can be used for the case \(n=1\) in Section \ref{sec:numcont}. This procedure is equivalent to running Hamiltonian dynamics on a constraint curve, which suggests a general approach for sampling based on (deterministic) Hamiltonian dynamics for \(n\geq 1\). We present the direct extension to multidimensional constraint manifolds in Section \ref{sec:constrainedMD} and then explicitly consider its application to the Bayesian framework with a specific (stochastic) numerical integrator in Section \ref{sec:hmc}. 

The resulting method is related to ones used to constrain the lengths and angles of bonds in molecular simulations \cite{Andersen_1983, Barth_1995, Lee_2005, Leimkuhler_2016}, which enables larger timesteps. More generally, \cite{Xu_2024} shows that constrained distributions involving up to \(\sim\)\(10^3\) variables can be tractably sampled by a symmetric quasi-Newton method similar to the approach to solve for constraints that we describe in Section \ref{sec:nonlinear_solve}. Considerations for convergence and bias of constrained Monte Carlo methods are discussed in \cite{Lelievre_2019, Lelievre_2022}.

\subsection{Numerical continuation in a single parameter}\label{sec:numcont}

\begin{figure}[t]
        \centering
        \includegraphics[width=\textwidth]{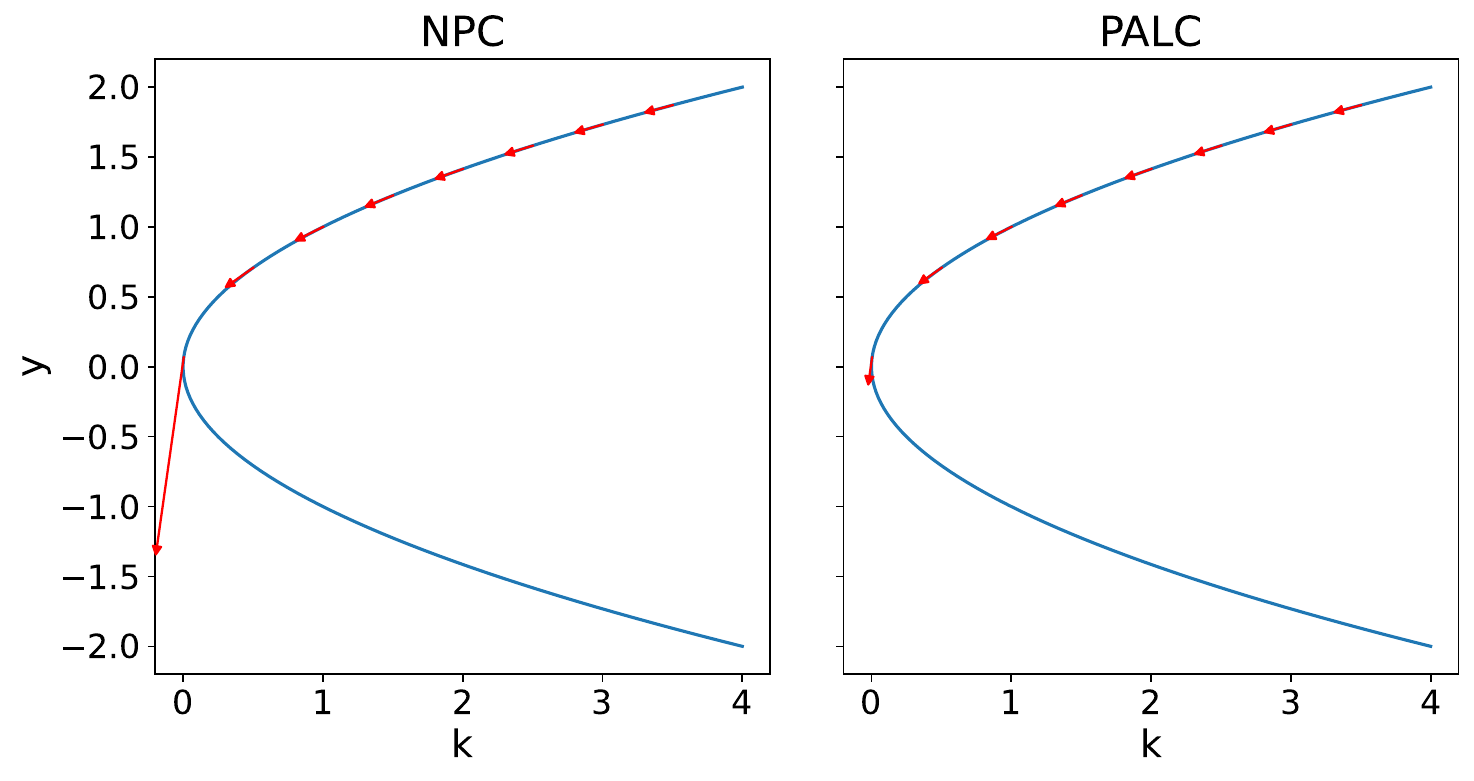}
       
    \caption{
     Behavior of the derivative vector \(\left(\begin{array}{cc}\dot{\mathbf y}&\dot{\mathbf k}\end{array}\right)\) when approaching a turning point at \((\mathbf 0, \mathbf 0)\). (left) In the case of natural parameter continuation, the derivative vector of (\ref{eqn:npc}) increases in norm when \((\mathbf y, \mathbf k)\) approaches \((\mathbf 0, \mathbf 0)\) due to the constraint \(\dot{\mathbf k}(t) = 1\) and is undefined at \((\mathbf 0, \mathbf 0)\). (right) In pseudoarclength continuation (PALC), the derivative vector is constrained to have unit norm in the system of equations in  (\ref{eqn:palc_dae}) and is well defined even at \((\mathbf y, \mathbf k) = (\mathbf 0, \mathbf 0)\).\label{fig:npc}
     \label{fig:palc}}
    \label{fig:continuation_fold}
\end{figure}

We first consider the case where \(\mathbf f:\mathbb R^{d+1}\rightarrow\mathbb R^d\), \(\mathbf y\in\mathbb R^d\), \(\mathbf k\in \mathbb R\). 
Suppose that \((\mathbf y(0),\mathbf k(0))\) is a solution to \(\mathbf f(\mathbf y, \mathbf k) = \mathbf 0\) and let  \(\mathbf f_\mathbf y\) denote the matrix with elements $\partial f_i/\partial y_j$. If \(\mathbf f_\mathbf y\) is nonsingular at \((\mathbf y(0),\mathbf k(0))\), then by the implicit function theorem, there is a one-dimensional curve \(\mathcal M\subset \mathbb{R}^{d+1} \) in a neighborhood of \((\mathbf y(0), \mathbf k(0))\) consisting of solutions to \(\mathbf f(\mathbf y,\mathbf k) = \mathbf 0\). To obtain additional solutions to \(\mathbf f(\mathbf y,\mathbf k) = \mathbf 0\), \(\mathcal M\) can be parameterized as the \(t\)-dependent solution to \(\mathbf f(\mathbf y(t), \mathbf k(t)) = \mathbf 0\), and a path of solutions that originate from the initial known solution \((\mathbf y(0), \mathbf k(0))\) can be extended as long as \(\mathbf f_\mathbf y\) remains nonsingular. 
A simple way to parameterize \(\mathcal M\) is
\begin{equation}
    \begin{aligned}
        \mathbf f(\mathbf y(t), \mathbf k(t)) &= \mathbf 0\\
        \dot{\mathbf k}(t) &= \mathbf{1},
    \end{aligned}
\end{equation}
where \(\dot{\mathbf k}\) is the derivative of \(\mathbf k\) with respect to \(t\), and $\mathbf 0$ and $\mathbf 1$ are vectors of zeros and ones, respectively. This parameterization is known as natural parameter continuation (NPC). Differentiating \(\mathbf f(\mathbf y(t), \mathbf k(t))\) results in the system of ODEs
\begin{subequations}\label{eqn:npc}
    \begin{align}
        \dot{\mathbf y}(t) &= -\mathbf f_{\mathbf y}(\mathbf y(t), \mathbf k(t))^{-1}\mathbf f_{\mathbf k}(\mathbf y(t), \mathbf k(t))\label{eqn:npc_tangent}\\
        \dot {\mathbf k}(t) &= \mathbf 1.
    \end{align}
\end{subequations}
In principle, the curve defined by \(\mathbf f(\mathbf y, \mathbf k) = \mathbf 0\) can be computed by numerically integrating \eqref{eqn:npc}, but this approach generally leads to issues: solving \eqref{eqn:npc} requires inverting \(\mathbf f_{\mathbf y}(t)\), and the solution is undefined when \(\mathbf f_{\mathbf y}(t)\) is singular, which generically occurs at turning points in the solution curve of \(\mathbf f(\mathbf y, \mathbf k) = \mathbf 0\) (Figure \ref{fig:npc}(left)).
One way to resolve these issues is to parameterize \(\mathcal M\) instead as the solution to the equations
\begin{subequations}
    \begin{align}
        \mathbf f(\mathbf y(t), \mathbf k(t)) &= \mathbf 0\\
        \left\|\dot{\mathbf y}(t)\right\|^2 + \|\dot{\mathbf k}(t)\|^2 &= \mathbf 1.\label{eqn:palc_daekin}
    \end{align}\label{eqn:palc_dae}
\end{subequations}
This parameterization is used to define the pseudoarclength continuation (PALC) procedure in \cite{Keller_1982}. As previously, we can differentiate \(\mathbf f(\mathbf y(t), \mathbf k(t))\) with respect to \(t\) to obtain a system of ODEs that can be numerically integrated.
Equation \eqref{eqn:palc_daekin} has the form of a kinetic energy, so (\ref{eqn:palc_dae}) can be interpreted as describing the motion of a particle constrained to slide along a one-dimensional rail at constant speed. Because the speed is constant, \eqref{eqn:palc_dae} is well-behaved even when \(\mathbf f_{\mathbf y}\) is singular (Figure \ref{fig:palc}(right)). This suggests constrained Hamiltonian dynamics as a natural generalization for computing points on constraint manifolds in problems involving more than one parameter.

\subsection{Constrained Hamiltonian dynamics} \label{sec:constrainedMD}

%


For clarity, we now treat the dynamical variables and the parameters together by defining \(\mathbf q \equiv \left(\begin{array}{cc}\mathbf y&\mathbf k\end{array}\right)\). 
Letting \(\mathbf p\) denote the conjugate momenta, the Hamiltonian of a general mechanical system is 
\begin{equation}
\mathcal H(\mathbf q,\mathbf p) = T(\mathbf p) + U(\mathbf q),
\end{equation}
where \(T(\mathbf p) = \frac{1}{2}\mathbf p^\mathsf{T}\mathbf M^{-1}\mathbf p\) is the kinetic energy, $U(\mathbf q)$ is the potential,  and \(\mathbf M\) is a symmetric positive definite mass matrix. The constrained dynamics of the system can be obtained by solving
\begin{equation}
\begin{aligned}
\dot{\mathbf q} &= \mathcal H_{\mathbf p}(\mathbf q,\mathbf p)= \mathbf M^{-1}\mathbf p\\
\dot{\mathbf p} &= -\mathcal H_\mathbf{q}(\mathbf q,\mathbf p)- \boldsymbol\lambda^\mathsf{T}\mathbf c_\mathbf{q}(\mathbf q)= -U_\mathbf{q}(\mathbf q) - \boldsymbol\lambda^\mathsf{T}\mathbf c_\mathbf{q}(\mathbf q)\\
\mathbf 0 &= \mathbf c(\mathbf q),
\end{aligned}\label{eqn:hamilton_eqs}
\end{equation}
where \(U_\mathbf q\) is the gradient of $U$,  \(\mathbf c_\mathbf q\) denotes the matrix with elements $\partial c_i/\partial q_j$, and \(\boldsymbol\lambda\in\mathbb R^m\) is a vector of Lagrange multipliers.
In general, (\ref{eqn:hamilton_eqs}) cannot be solved analytically and must be integrated numerically. We use an operator splitting method \cite{Leimkuhler_2016} to integrate (\ref{eqn:hamilton_eqs}), as we describe in Section \ref{sec:split}.

\subsection{Hamiltonian Monte Carlo methods}\label{sec:hmc}

Our goal is to sample a distribution,  \(\pi(\mathbf q)\) (e.g., representing \eqref{eq:bayes}), subject to constraints (e.g., imposed by \eqref{eqn:palc_dae} for an $\mathbf f$ such as that in \eqref{eqn:repressilator}).   To this end, we define the potential
\begin{equation}
U(\mathbf q) = -\log(\pi(\mathbf q)).
\end{equation}
Deterministically integrating (\ref{eqn:hamilton_eqs}) does not generate samples from the whole target distribution owing to energy conservation, and it is necessary to introduce randomness. A few schemes with empirically good performance are jittered Hamiltonian Monte Carlo \cite{Hoffman2021AnAS}, no-U-turn sampling \cite{NUTS}, and underdamped Langevin dynamics \cite{rioudurand2022metropolis}. The advantage of these schemes over RWM is that they make use of the gradient of $U(\mathbf q)$, which directs sampling toward low-energy/high-probability states. For certain classes of target distributions, such as those consisting of \(n\) independent and identically distributed components \cite{Beskos_2013} and strongly log-concave distributions that are sufficiently smooth \cite{Mangoubi_2018}, the advantage of gradient based samplers over RWM can be shown to grow with dimension: \(\mathcal O(n^{1/4})\) steps are required to generate an independent sample  using Hamiltonian Monte Carlo, whereas \(\mathcal O(n)\) steps are required when using RWM.

For the present study, we use Langevin dynamics.  To this end, we modify (\ref{eqn:hamilton_eqs}) to obtain the system of stochastic differential equations:
\begin{subequations}\label{eqn:langevin}
    \begin{align}
    \dot{\mathbf q} &= \mathbf M^{-1}\mathbf p\\
    \dot{\mathbf p} &= -U_\mathbf{q} - \gamma\mathbf p + \sqrt{2T\gamma}\mathbf M^{1/2}\boldsymbol{\eta}(t) - \boldsymbol\lambda^\mathsf{T}\mathbf c_\mathbf{q}(\mathbf q)\\
    \mathbf 0 &= \mathbf c(\mathbf q),
    \end{align}
\end{subequations}
where \(\gamma\) is a friction coefficient, \(T\) is the temperature, and \(\boldsymbol \eta\) is a white-noise vector of independent components, each with unit variance. We integrate these equations numerically by building on the operator splitting scheme mentioned previously \cite{Leimkuhler_2016}, as we describe in Section \ref{sec:obabo}. 

Our method for enforcing constraints is related to the SHAKE \cite{Ryckaert_1977} and RATTLE \cite{Andersen_1983} methods used in molecular simulation.  SHAKE with the velocity Verlet integrator can be viewed as an unconstrained half step in \(\mathbf p\), followed by a step in \(\mathbf q\) with projection onto the constraint manifold \(\mathcal M\), and then another unconstrained half step in \(\mathbf p\). However, even if the position constraints are satisfied, the dynamics with SHAKE do not satisfy the cotangency constraints \(\mathbf c_{\mathbf q}(\mathbf q)\mathbf M^{-1}\mathbf p = \mathbf 0\) in general. To enforce the cotangency constraints, RATTLE with the velocity Verlet integrator adds an orthogonal projection of \(\mathbf p\) onto the cotangent space after the last half step in \(\mathbf p\). The operator splitting method that we use (see Section \ref{sec:obabo}) is similar to RATTLE with the velocity Verlet integrator but also performs an orthogonal projection of \(\mathbf p\) onto the cotangent space after the first half step in \(\mathbf p\).

Because the nonlinear projection performed in our sampler is the same as that performed in pseudoarclength continuation, our sampler inherits the properties of pseudoarclength continuation; in particular, the dynamics can move past fold bifurcations if the step size is small enough. However, other bifurcations may require special treatment. One can monitor and control for local bifurcations by including auxiliary variables and constraint equations that define the bifurcation during sampling, as we do for Hopf bifurcations in Section \ref{sec:hopf_points}.

\section{Estimating model parameters for the repressilator}\label{sec:calculations}

To demonstrate our method and compare its performance with established Monte Carlo methods for Bayesian parameter estimation, we fit the model for the repressilator in \eqref{eqn:repressilator} to simulated data. When the number of species in the cycle is odd (integer $\ell$ in \eqref{eqn:repressilator} and Figure \ref{fig:repressilator_diagram}), the dynamics tend to either a globally attracting fixed point or a globally attracting limit cycle \cite{Smith_1987}. These relatively simple dynamics make the repressilator a tractable yet nontrivial test of our methods.

To enforce nonnegativity of both concentrations and rate constants, we make a change of variables to logarithmic coordinates:
\begin{equation}
    \begin{aligned}
        \left(\begin{array}{cc}\tilde{\mathbf y}& \tilde{\mathbf k}\end{array}\right)&\equiv\left(\begin{array}{cc}\log(\mathbf y)& \log(\mathbf k)\end{array}\right)\\
        \frac{d\tilde{\mathbf y}}{ds} &= \tilde{\mathbf f}(\tilde{\mathbf y}, \tilde{\mathbf k}, \mathbf n)\\
        f_j(\tilde{\mathbf y}, \tilde{\mathbf k}, \mathbf n) &= \frac{\exp\left(\tilde{k}_{j,0} - \tilde{y}_j\right)}{1 + \exp\left(n_{j - 1}\tilde{y}_{j - 1}\right)} - \exp\left(\tilde{k}_{j,1} - \tilde{k}_{0,1}\right),~j\in\{0,1,\dots,2l\}.\label{eqn:repressilator_log}
    \end{aligned}
\end{equation}
For notational convenience, we henceforth omit the tildes on \(\tilde{\mathbf y}\) and \(\tilde{\mathbf k}\) where there is no risk of confusion. The repressilator model with \(2\ell + 1\) species has a total of \(6\ell + 2\) independent parameters after setting the timescale by fixing \({k}_{0,1} = 0\): the synthesis rates \(\{k_{0, i}\}_{i=0}^{2\ell}\), the remaining degradation rates \(\{k_{1, i}\}_{i=1}^{2\ell}\), and the Hill coefficients \(\{n_i\}_{i=0}^{2\ell}\). We generate simulated data by choosing arbitrary values for the parameters, integrating the equations forward in time using Tsitouras's 5/4 explicit Runge-Kutta method \cite{Tsitouras_2011}, and adding normally distributed noise with mean zero and variance \(2.5\times10^{-3}\) to the concentration of the first species (prior to transforming to logarithmic coordinates) (Figure \ref{fig:repressilator_data}).  That is, we assume only one of the $2\ell +1$ species is observable.
We then sample from a sequence of constrained distributions, each of which incorporates more information regarding the model and is more computationally demanding than the previous one. 

The results are summarized in Figure \ref{fig:hist_rp}, where we project the sampled parameter sets onto three pairs of parameters. The top row shows sampling of all fixed points.  The sampling is nearly uniform, indicating that the procedure readily explores the space. The middle row shows sampling of Hopf bifurcations.  The constraints on the Jacobian (discussed below) restrict the sampling to a subspace, though it remains quite smooth. The bottom row shows sampling of limit cycles that fit the data.  The sampling is most constrained in this case.  We discuss how we obtain these results in Sections \ref{sec:fixed_points} to \ref{sec:limit_cycles}, and we compare our procedure with unconstrained MCMC algorithms in Section \ref{sec:MCMCcomp}. We use a step size of \(10^{-1}\), friction coefficient \(\gamma = 10^{-1}\), and mass matrix \(\mathbf M = \mathbf I\) throughout.

\subsection{Sampling fixed points}\label{sec:fixed_points}
Let \(\mathbf q \equiv \left(\begin{array}{ccc}\mathbf y&\mathbf k&\mathbf n\end{array}\right)\). If the right hand side of (\ref{eqn:repressilator_log}) is specified as the constraint \(\mathbf c(\mathbf q)=\mathbf{0}\) to be satisfied in (\ref{eqn:hamilton_eqs}), then assuming that \(\mathbf c_{\mathbf q}\) is nonsingular, every \(\mathbf q\) sampled gives a set of parameters for which (\ref{eqn:repressilator_log}) has a fixed point at \({\mathbf y}\). This can be useful, for instance, if the steady-state behavior of an experimental system is known, allowing the space of parameters to be restricted without performing costly integration of the system of ODEs.

During sampling, we bound the search space by including half quadratic penalty terms in the log-prior:
\begin{equation}
    \begin{aligned}
    100\left\|(\mathbf k - \mathbf k_{\text{max}})\odot\mathbbm 1\{\mathbf k > \mathbf k_{\text{max}}\}\right\|^2 &+
    100\left\|(\mathbf k - \mathbf k_{\text{min}})\odot\mathbbm 1\{\mathbf k < \mathbf k_{\text{min}}\}\right\|^2 +\\
    100\left\|(\mathbf n - \mathbf n_{\text{max}})\odot\mathbbm 1\{\mathbf n > \mathbf n_{\text{max}}\}\right\|^2&+
    100\left\|(\mathbf n - \mathbf n_{\text{min}})\odot\mathbbm 1\{\mathbf n < \mathbf n_{\text{min}}\}\right\|^2,
   \end{aligned}
    \label{eqn:box_restraints}
\end{equation}
where \(\mathbbm1\{\mathbf k > \mathbf k_\text{max}\}\) compares \(\mathbf k\) and \(\mathbf k_\text{max}\) componentwise and returns a vector with components equal to \(1\) wherever \(\mathbf k > \mathbf k_\text{max}\) and 0 otherwise, and \(\odot\) denotes the componentwise product.  We choose \(\mathbf k_\text{min}\) to be a length \(4\ell + 1\) vector with all components equal to \(-5\), \(\mathbf k_\text{max}\) to be a length \(4\ell + 1\) vector with all components equal to \(5\), \(\mathbf n_\text{min}\) to be a length \(2\ell + 1\) vector with all components equal to \(0\), and \(\mathbf n_\text{max}\) to be a length \(2\ell + 1\) vector with all components equal to \(10\). We sample a single chain for \(10^7\) steps and subsample every 10 steps. 

\begin{figure}[h]
    \centering
    \includegraphics[width=\textwidth]{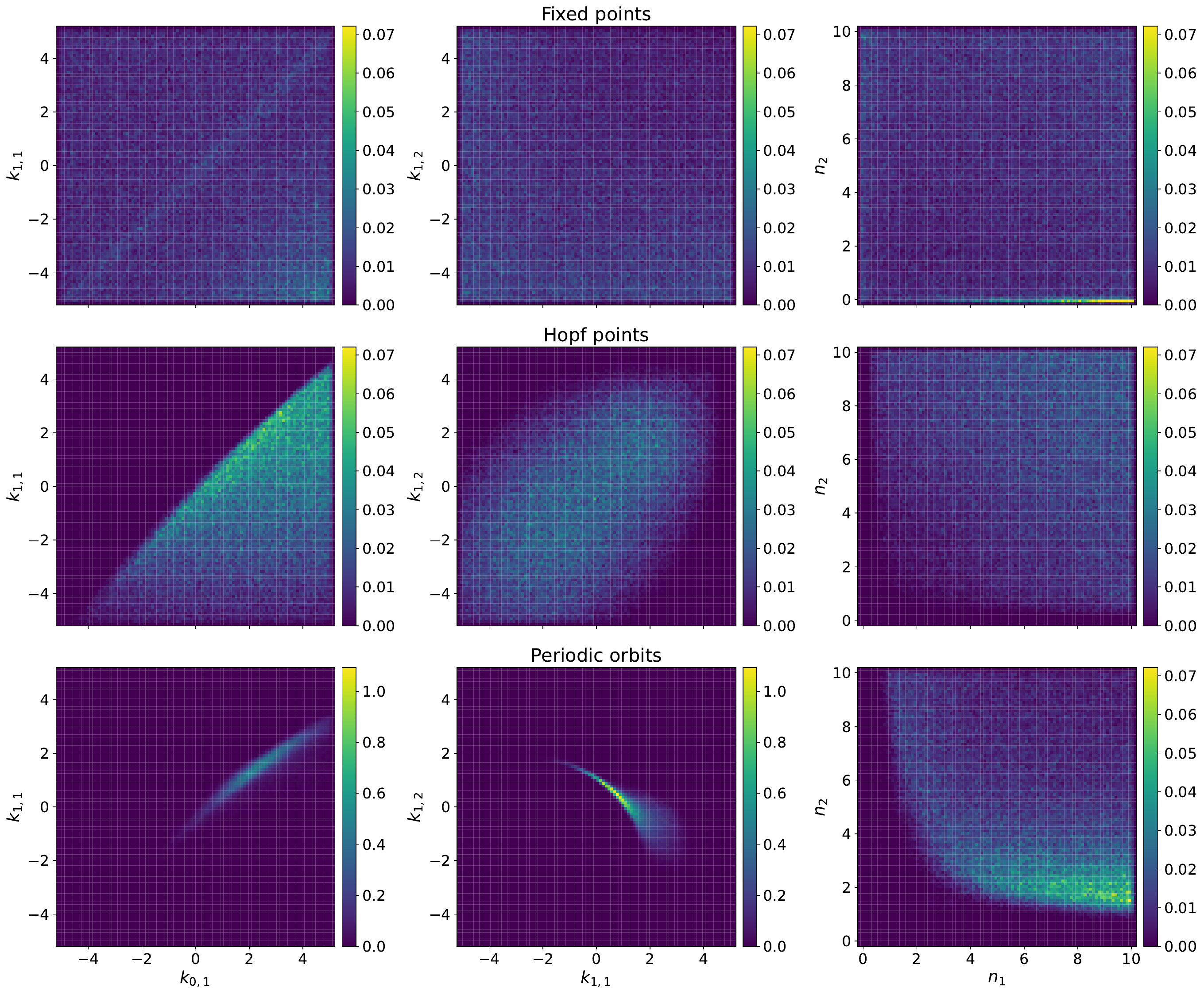}
    \caption{Representative two-dimensional marginals for the three-species represillator model parameter distribution obtained from \(10^{7}\) steps of the constrained Langevin sampler with fixed-point constraints (top), Hopf-point constraints (middle), and periodicity constraints (bottom). The log-likelihood when sampling fixed points and Hopf points is constant everywhere, and the log-likelihood when sampling limit cycles is proportional to the squared error from the data shown in Figure \ref{fig:repressilator_data} (see Section \ref{sec:MCMCcomp}).}
    \label{fig:hist_rp}
\end{figure}

As noted above, the results are shown in the top row of Figure \ref{fig:hist_rp}. Since no additional terms other than the restraints \eqref{eqn:box_restraints} are included in the posterior, the sampled points should be uniformly distributed across the solution set of \(\mathbf f(\mathbf y, \mathbf k, \mathbf n) = \mathbf 0\). However, we note that the points are not necessarily uniformly distributed throughout the parameter space due to the curvature of the solution manifold. In particular, the top right panel of Figure \ref{fig:hist_rp} shows points concentrated along \(n_2 = 0\). This is a consequence of sampling in logarithmic concentrations. When \(n_2\) is small, combinations of parameters that cause the steady-state concentration of \(y_1\) to be large correspondingly cause the steady-state concentration of \(y_2\) to be close to zero. The logarithm of the concentration of \(y_2\) in this case is uniformly distributed across a large range of negative values, which results in the sampled points concentrating near \(n_2 = 0\) when they are projected onto the \((n_1, n_2)\) plane. If desired, one can correct for the curvature of the solution manifold when projecting onto the parameter space by reweighting by the transformed volume element of the solution manifold (see Section \ref{sec:reweight_curvature}).

\subsection{Sampling Hopf points}\label{sec:hopf_points}
For the repressilator, we are interested in estimating model parameters from oscillatory data, so we seek sets of parameter values that give rise to limit cycles. Empirically, we find that these sets represent a small fraction of the space of possible parameter values.  To boost efficiency, we can exploit the fact that one way for limit cycles to appear is through Hopf bifurcations and restrict the sampling to a manifold of Hopf bifurcation points. At a Hopf bifurcation, the Jacobian of a system of ODEs evaluated at a fixed point has a complex conjugate pair of imaginary eigenvalues that crosses the imaginary axis, causing the fixed point to change stability and give rise to a small amplitude limit cycle. 

Numerically, the exact point of crossing, at which the Jacobian has a purely imaginary pair of eigenvalues, can be computed as the solution to the system of equations
\begin{subequations}
    \begin{align}
        \mathbf f(\mathbf y, \mathbf k, \mathbf n) &= \mathbf 0\\
        (\mathbf f_{\mathbf y}(\mathbf y, \mathbf k, \mathbf n) - i\omega \mathbf I)\mathbf v &= \mathbf 0\\
        \mathbf v^*\mathbf v &= 1\\
        \operatorname{Im}(v_1) &= 0,
    \end{align}\label{eqn:hopf_fully_extended_complex}
\end{subequations}
where \(\mathbf v\in\mathbb C^{d}\) is the eigenvector of \(\mathbf f_\mathbf y(\mathbf y,\mathbf k)\) associated with the eigenvalue \(i\omega\), and \(v_1\) denotes the first component of \(\mathbf v\). We incorporate \eqref{eqn:hopf_fully_extended_complex} as real valued constraints to be satisfied during sampling after splitting \(\mathbf v\) into its real and imaginary parts:

    \begin{equation}
        \mathbf c({\mathbf q}) \equiv \left\{\begin{aligned}
                \mathbf f(\mathbf y, \mathbf k, \mathbf n) &= \mathbf 0\\
                \mathbf f_{\mathbf y}(\mathbf y, \mathbf k, \mathbf n)\operatorname{Re}(\mathbf v) + \omega\operatorname{Im}(\mathbf v) &= \mathbf 0\\
                \mathbf f_{\mathbf y}(\mathbf y, \mathbf k, \mathbf n)\operatorname{Im}(\mathbf v) - \omega\operatorname{Re}(\mathbf v) &= \mathbf 0\\
                \operatorname{Re}(\mathbf v)^{\mathsf T}\operatorname{Re}(\mathbf v) + \operatorname{Im}(\mathbf v)^{\mathsf T}\operatorname{Im}(\mathbf v) &= 1\\
                \operatorname{Im}(v_1) &= 0.
            \label{eqn:hopf_fully_extended}
        \end{aligned}\right.
    \end{equation}

This is the fully extended system for Hopf bifurcations \cite{Govaerts_2000}. Sampling \(\mathbf q\equiv \left(\begin{array}{ccccc} \mathbf y&\mathbf k&\mathbf n&\mathbf v&\omega \end{array}\right)\) subject to \eqref{eqn:hopf_fully_extended} gives \(\mathbf k\) and \(\mathbf n\) for which \(\mathbf f(\mathbf y,\mathbf k, \mathbf n)\) has a Hopf point at \(\mathbf y\). Solving \eqref{eqn:hopf_fully_extended} is computationally more demanding than solving for the fixed points of \eqref{eqn:repressilator_log} but is still orders of magnitude computationally cheaper than numerically integrating \eqref{eqn:repressilator_log}. The period estimate \(2\pi/\omega\) can be used to  restrict the space of parameters for ODE models of oscillatory systems further, and the results sampling with Hopf point constraints provide reasonable starting points for performing sampling that involves the computationally expensive time integration of (\ref{eqn:repressilator_log}).

The fixed points of  (\ref{eqn:repressilator_log}) sampled in Section \ref{sec:fixed_points} cannot be directly used to initiate Hopf-point sampling since they generally do not have any purely imaginary eigenvalues and thus do not satisfy \eqref{eqn:hopf_fully_extended}. To find starting points consistent with \eqref{eqn:hopf_fully_extended}, we compute the eigenvalues and eigenvectors of the Jacobian matrix at each of the fixed points sampled in Section \ref{sec:fixed_points} and then select the eigenvalue-eigenvector pair whose eigenvalue has the smallest real part relative to its imaginary part: 
\begin{equation}
    \zeta^*_i = \min_{j} \left|\frac{\operatorname{Re}(\zeta_{ij})}{\operatorname{Im}(\zeta_{ij})}\right|,
\end{equation}
where \(\zeta_{ij}\) is the \(j\)th eigenvalue of the \(i\)th fixed point sampled and \(\mathbf v_{ij}\) is its corresponding eigenvector. We then select \(\left(\begin{array}{ccc}\mathbf y_i& \mathbf k_i&\mathbf n_i\end{array}\right)\) for which \(|\operatorname{Re}(\zeta_{i}^*)/\operatorname{Im}(\zeta_{i}^*)|\) is small, set \(\mathbf q_0 \equiv \left(\begin{array}{ccccc}\mathbf y_i&\mathbf k_i&\mathbf n_i&\mathbf v_{i}^*&\mathbf \zeta_{i}^*\end{array}\right)\), and solve \eqref{eqn:hopf_fully_extended} using Gauss--Newton iteration starting from \(\mathbf q_0\) as an initial guess. Upon success, we obtain a point \(\mathbf q\) consistent with the constraints in \eqref{eqn:hopf_fully_extended} from which we initiate Hopf-point sampling.

During Hopf-point sampling, we again bound the search with the prior in \eqref{eqn:box_restraints}. We sample a single chain for \(10^7\) steps and subsample every 10 steps. The results are shown in the middle row of Figure \ref{fig:hist_rp}.


\subsection{Sampling limit cycles} \label{sec:limit_cycles}
With candidate parameters obtained from sampling the solution manifold of (\ref{eqn:hopf_fully_extended}), further sampling involving the explicit computation of limit cycle solutions to (\ref{eqn:repressilator_log}) can be initiated from selected points. We formulate the limit cycle computation as a boundary value problem: 
\begin{equation}
    \begin{aligned}
        \frac{d \mathbf y}{ds} &= \mathbf f(\mathbf y, \mathbf k)\\
        \mathbf y(0) &= \mathbf y(\tau).
    \end{aligned}\label{eqn:periodic_bvp}
\end{equation}
We discretize (\ref{eqn:periodic_bvp}) using fourth-order Gauss--Legendre collocation on 60 mesh intervals (see Section \ref{sec:collocation}) to obtain a system of \(241(2\ell + 1)\) equations whose solution \(\mathbf u(s)\) approximates \(\mathbf y(s)\). Incorporating the discretized equations into \(\mathbf c(\mathbf q)\) and sampling \(\mathbf q = \left(\begin{array}{cccc} \mathbf u(s)&\mathbf k&\mathbf n&\tau \end{array}\right)\) gives points that allow for periodic solutions to (\ref{eqn:repressilator_log}). While this constraint does not exclude fixed point solutions, which are trivially periodic, such solutions can be discouraged by including a penalty term for solutions with arclengths below a threshold in the prior. 
We compute the arclength as
\begin{equation}
        L = \int_{0}^{\tau}\|\dot{\mathbf u}(s)\|\mathop{ds}
\end{equation}
and use a penalty of the form
\begin{equation}
        \left(\left(\frac{L_0}{L\sqrt{2}}\right)^{4} - \left(\frac{L_0}{L\sqrt{2}}\right)^{2} + \frac{1}{4}\right)\mathbbm1\{L < L_0\}
    \label{eqn:arclength_prior}
\end{equation}
with $L_0 = 0.3$ for the results shown.

In addition to the prior terms in  \eqref{eqn:box_restraints} and \eqref{eqn:arclength_prior}, we include a least squares log-likelihood term of the form in \eqref{eqn:LL_lstsq}. 
Thus the full log-posterior is
\begin{equation}
    \begin{aligned} \pi (\mathbf q|\mathbf x) &= \sum _{i=1}^N\sum _{j=1}^{M_i}\frac {1}{2\sigma _{i}^{2}}\left (\hat x_i - x_i\right )^2 \\ 
    &\quad +\left (\left (\frac {L_0}{L\sqrt {2}}\right )^{4} - \left (\frac {L_0}{L\sqrt {2}}\right )^{2} + \frac {1}{4}\right )\mathbbm{1}\{L < L_0\} +\frac {\left (\tau - \tau _{\mathrm{data}}\right )^2}{2\sigma^2}\\ 
    &\quad+100\left \|(\mathbf k - \mathbf k_{\text{max}})\odot \mathbbm{1}\{\mathbf k > \mathbf k_{\text{max}}\}\right \|^2+100\left \|(\mathbf k - \mathbf k_{\text{min}})\odot \mathbbm{1}\{\mathbf k < \mathbf k_{\text{min}}\}\right \|^2\\ 
    &\quad+100\left \|(\mathbf n - \mathbf n_{\text{max}})\odot \mathbbm{1}\{\mathbf n > \mathbf n_{\text{max}}\}\right \|^2+100\left \|(\mathbf n - \mathbf n_{\text{min}})\odot \mathbbm{1}\{\mathbf n < \mathbf n_{\text{min}}\}\right \|^2. \end{aligned}
\end{equation}

We process the simulated data shown in Figure \ref{fig:repressilator_data} to use as \(\mathbf x\) in the log-likelihood. We estimate the period \(\tau_\mathrm{data}\) by dividing the total integration time by the highest amplitude frequency of the fast Fourier transform of the data, excluding the zero-frequency mode. We then segment the data into pieces of length \(\tau_\mathrm{data}\), average them, and scale the time so the period is 1. The model output \(\hat{\mathbf x}\) in the least squares log-likelihood is \(\exp(y_0)\) evaluated at the time associated with each data point. We use \(\sigma_i=0.05\) for all $i$ in the log-likelihood term \eqref{eqn:LL_lstsq}.  We reintroduce the period by including a quadratic penalty on the difference between the period of the model output \(\tau\) and \(\tau_\mathrm{data}\) in the log-likelihood: $\left(\tau - \tau_\mathrm{data}\right)^2/2\sigma^2$ with $\sigma=0.05$.  

To initiate sampling, we choose \(\mathbf k\) from the Hopf-point samples that minimize \((2\pi/\omega - \tau_\mathrm{data})^2\) and use a fifth-order Radau IIA method to integrate \eqref{eqn:repressilator_log} forward starting from the initial conditions that were used to generate the simulated data (which we assume are known) for \(20\pi/\omega\) time units. The solution over the last \(2\pi/\omega\) length interval of time is taken to be \(\mathbf y(t)\). This \(\mathbf y(t)\), along with \(\mathbf k\) and \(\tau\), is used as an initial guess to solve \eqref{eqn:periodic_bvp} using Gauss--Newton iteration to obtain an initial point for sampling with periodicity constraints. We sample 10 independent chains for \(10^7\) steps each and subsample every 10 steps. Results are shown in the bottom row of Figure \ref{fig:hist_rp}.

\subsection{Comparison with unconstrained MCMC methods}\label{sec:MCMCcomp}

To assess the speedup from our method, we compare our results to three widely-used MCMC methods: an affine-invariant ensemble sampler (henceforth, ensemble MCMC) \cite{Goodman_2010, Foreman_Mackey_2013}, the unconstrained Metropolis-adjusted underdamped (i.e., intertial) Langevin algorithm (MALA) \cite{Horowitz_1991}, and the no-U-turn sampler (NUTS) \cite{NUTS}. Descriptions of each of the methods can be found in Section \ref{sec:unconstrained_mcmc}.

As we do in Section \ref{sec:limit_cycles}, for the constrained Langevin dynamics, we use the restraint in \eqref{eqn:box_restraints} to bound the search space, the arc-length penalty term in \eqref{eqn:arclength_prior} to discourage constant solutions, and a log-likelihood of the form \eqref{eqn:LL_lstsq}. In addition to the model parameters \(\mathbf k\) and \(\mathbf n\), we include the initial conditions \(\mathbf y(0)\) as parameters to be sampled to align the phase of the model output with the phase of the data. The total set of parameters sampled with the unconstrained methods is \(\left(\begin{array}{ccc}\mathbf k&\mathbf n&\mathbf y(0)\end{array}\right)\). We process the simulated data the same way as described in Section \ref{sec:limit_cycles} to obtain \(\mathbf x\) but instead average over a window of \(2\tau_\mathrm{data}\) to better control the period of the model output. To obtain the model output \(\hat{\mathbf x}\) when evaluating the log-likelihood for the unconstrained MCMC methods, we use Tsitouras's 5/4 explicit Runge-Kutta method \cite{Tsitouras_2011} to integrate \eqref{eqn:repressilator_log} forward for 30 time units starting from \(\mathbf y(0)\). We then take \(\hat{\mathbf x}\) to be \(\exp(y_0)\) on the interval \(t \in [30 - 2\tau_\mathrm{data}, 30]\). 

For all the methods, the bulk of the computational cost comes from numerical integration of \eqref{eqn:repressilator_log}: for the CMALA and CULA, we solve a two-point boundary value problem for the constraints at each step, and, for the unconstrained methods, we solve an initial value problem at each step (each attempted walker update in ensemble MCMC and each iteration of the symplectic integrator in MALA and NUTS). 
As such, the cost of a step is similar for all the methods, and we compare their performance in terms of the total number of steps.

\begin{table}
    \centering
    \begin{tabulary}{\textwidth}{|c|c|C|C|C|C|}
        \hline
         Species&Method&{Acceptance rate}&{Steps for \(\hat{R} < 1.1\)}&{Average ESS/step} & {Minimum ESS/step}\\
         \hline\hline
         \rowcolor{gray!20}
         3&EMCEE &0.0397&\(2\times10^{8}\)&\(9.96\times10^{-6}\)&\(4.55\times10^{-6}\)\\\hline
         3&CMALA&0.8332&\(1\times10^{7}\)&\(1.87\times10^{-4}\)&\(8.61\times10^{-5}\)\\\hline
         \rowcolor{gray!20}
         3&CULA&0.9561&\(1\times10^{6}\)&\(2.77\times10^{-3}\)&\(1.76\times10^{-3}\)\\\hline
         3&MALA&0.6277&\(\gg1\times10^8\)&\(6.95\times10^{-8}\)&\(6.19\times10^{-8}\)\\\hline
         \rowcolor{gray!20}
         3&NUTS&0.8537&\(2.5\times10^{7}\)&\(4.12\times10^{-5}\)&\(1.87\times10^{-5}\)\\\hline
         7&EMCEE&0.0134&\(\gg4\times10^8\)&\(5.71\times10^{-7}\)&\(4.81\times10^{-8}\)\\\hline
         \rowcolor{gray!20}
         7&CMALA&0.7916&\(1.4\times10^{8}\)&\(9.23\times10^{-6}\)&\(9.27\times10^{-7}\)\\\hline
         7&CULA&0.9647&\(1.5\times10^{7}\)&\(1.38\times10^{-4}\)&\(1.28\times10^{-5}\)\\\hline
         \rowcolor{gray!20}
         7&NUTS&0.9087&\(5.4\times10^{7}\)&\(1.05\times10^{-5}\)&\(6.83\times10^{-6}\)\\\hline
    \end{tabulary}
    \caption{Comparison of the performance of constrained Metropolis-adjusted Langevin algorithm (CMALA), constrained unadjusted Langevin algorithm (CULA), unconstrained Metropolis-adjusted Langevin algorithm (MALA), no-U-turn sampler (NUTS), and ensemble MCMC (EMCEE) when estimating parameters for the three- and seven-species repressilator models using data that are consistent with those shown in Figure \ref{fig:repressilator_data}. The acceptance rate for CULA dynamics is below one because steps that fail to satisfy the constraints are rejected.  MALA was not tested for the seven-species model owing to its failure to converge for the three-species model.}
    \label{tab:performance_metrics}
\end{table}

One way to quantify the relative performance of the algorithms is the effective sample size (ESS) per step (Table \ref{tab:performance_metrics}). 
For a one-dimensional distribution, the ESS is given by
\begin{equation}
    N_\textrm{eff} = \frac{N}{-1 + 2\sum_{t=0}^\infty C(t)},
\end{equation}
where \(N\) is the number of samples, and \(C(t)\) is the autocorrelation function at time lag \(t\) \cite{Sokal_1997}. In practice, the sum must be truncated at some finite time lag. We use the criterion given in \cite{Geyer_1992}. For the example that we consider, there are multiple parameters, and we report the average and minimum ESS over all of them. 

To monitor convergence, we compute a version of the potential scale reduction factor \(\hat{R}\) described in \cite{Gelman_1992} for \(\mathbf k\). For a set of \(M\) independent chains each with \(N\) samples, $\hat{R}= \|\boldsymbol\Sigma_{a}^{-1}\boldsymbol\Sigma\|_2$, where $\boldsymbol\Sigma_{a}$ is the average within-chain covariance and $\boldsymbol\Sigma$ is an estimate of the covariance of the stationary distribution, which includes the between-chain covariance (see Section \ref{sec:GelmanRubin} for formulas).
If the sampling is converged, the within-chain covariance \(\boldsymbol\Sigma_a\) and stationary covariance \(\boldsymbol\Sigma\) matrices should be approximately equal, and the 2-norm (i.e., square root of the largest singular value) of the matrix \(\boldsymbol\Sigma_a^{-1}\boldsymbol\Sigma\) should be approximately one. 



\begin{figure}[bt]
    \centering
    \includegraphics[width=\textwidth]{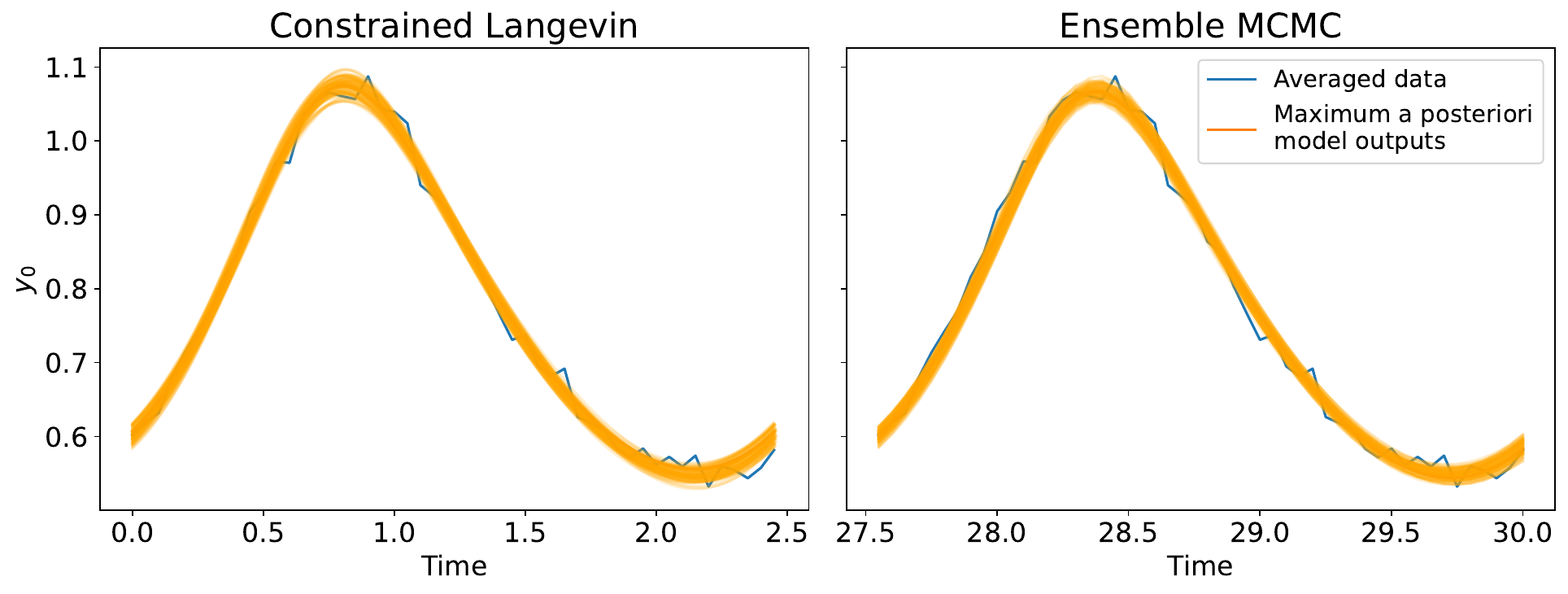}
    \caption{100 best fit solutions obtained after \(10^7\) steps of the CMALA sampler (left) and \(2\times10^8\) steps the ensemble MCMC sampler (right).  Results shown are for the three-species repressilator fit to the data shown in Figure \ref{fig:repressilator_data} (see Section \ref{sec:MCMCcomp}).}
    \label{fig:best_fit_comparison}
\end{figure}

\begin{figure}[hbt]
    \centering
    \includegraphics[width=\textwidth]{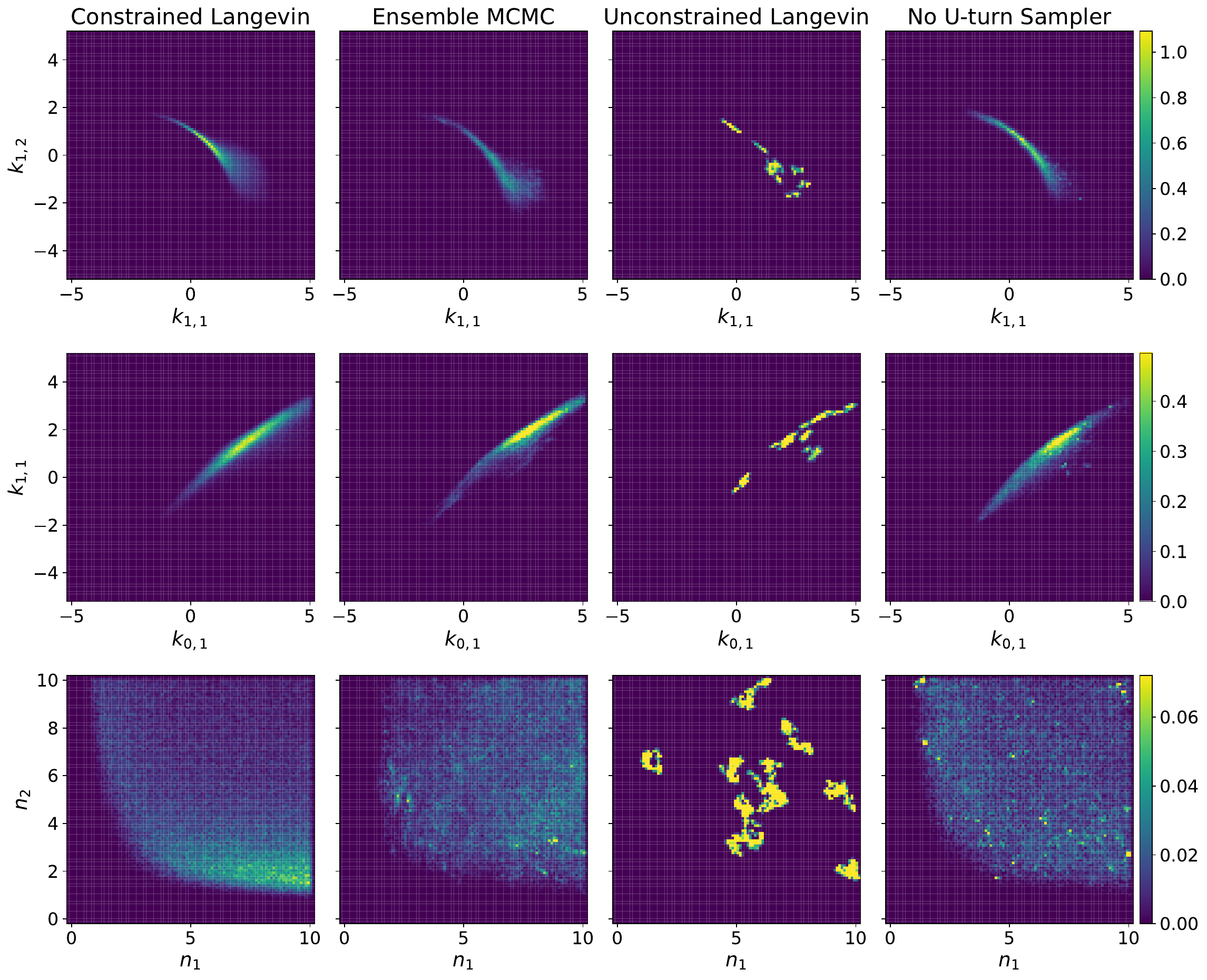}
    \caption{Representative marginal distributions sampled using constrained MALA, ensemble MCMC, unconstrained MALA, and NUTS after \(10^{7}\) steps. Results shown are for the three-species repressilator fit to the data shown in Figure \ref{fig:repressilator_data} (see Section \ref{sec:MCMCcomp}).}
    \label{fig:emcee_comparison}
\end{figure}

\begin{figure}
    \centering
    \centering
    \includegraphics[height=0.17\textheight]{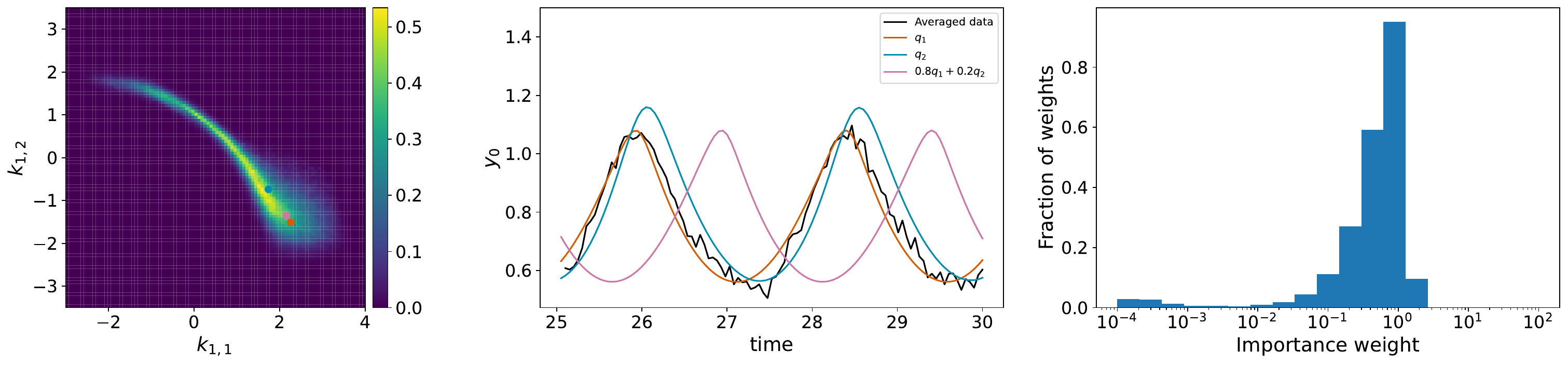}
    \includegraphics[height=0.17\textheight]{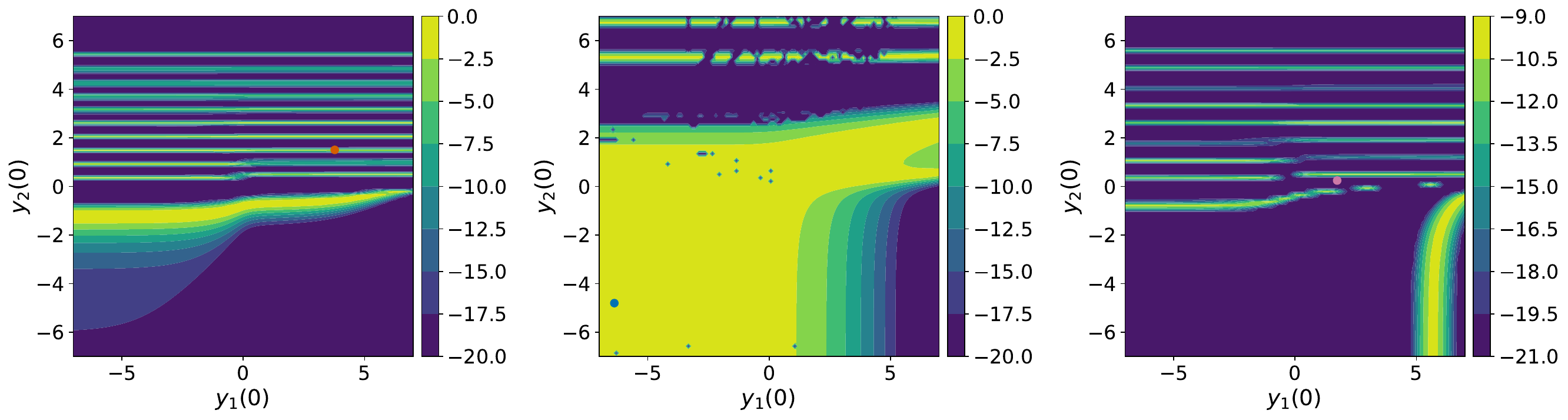}
    \caption{The distribution sampled by the unconstrained methods is multimodal despite the marginals along certain dimensions appearing smooth (top left). The multimodal structure is apparent when examining the distribution at different initial values of \(y_1\) and \(y_2\): the unnormalized log-posterior distribution exhibits rapid, non-concave variations with the rest of the parameters fixed at the high probability values indicated by the symbols (\(q_1\), orange; \(q_2\), blue; the linear interpolation \(0.8q_1 + 0.2q_2\); pink in the bottom row).  The interpolated parameters generate trajectories that are close in period and shape to the trajectories generated by \(q_1\) and \(q_2\) but about half a period out of phase with the data used to compute the log-likelihood (top middle). The apparent noise in the bottom middle plot results from failure to integrate the ODEs for certain initial conditions. Most of the normalized importance weights for reweighting the distribution sampled by constrained Langevin to the distribution sampled by the unconstrained methods are close to one (top right).}
    \label{fig:hist_emcee_interpolated}
\end{figure}

Table \ref{tab:performance_metrics} gives the ESS and potential scale reduction factors for each of the methods when applied to sampling parameters for the three- and seven-species repressilator models. Best fits and marginal distributions of the parameters for the three-species repressilator model are shown in Figures \ref{fig:best_fit_comparison} and \ref{fig:emcee_comparison}, respectively.
Both the constrained and unconstrained methods produce fits that are visually reasonable, but some unconstrained methods are less efficient at sampling. 
The ensemble MCMC sampler has a low acceptance rate due to the high sensitivity to changes in parameter values of initial-value problem solutions to \eqref{eqn:repressilator_log}: small changes in parameter values can cause large changes in the phase, resulting in rapid variations in the posterior distribution (Figure \ref{fig:hist_emcee_interpolated}). Consequently, proposals generated by \eqref{eqn:stretch_move} using two points close in Euclidean distance are frequently rejected as they often land in low probability regions. Consequently, the ESS per step of ensemble MCMC is about 20 times lower than that for constrained MALA (CMALA). Consistent with the ESS, the required number of steps for convergence is about 20 times higher for EMCEE than it is for CMALA.

Similarly, unconstrained MALA mixes extremely slowly; none of the walkers explore beyond the immediate vicinity of their starting points. The conditioning of the posterior requires MALA to use a step size 100 times smaller than the one that we use for CMALA to achieve a reasonable acceptance rate, making the ESS per step 2000 times lower for MALA than for CMALA and preventing convergence within the time that we allow (\(10^8\) steps). By solving for the periodic solutions directly, the constrained Langevin sampler is agnostic to initial conditions, resulting in a much smoother posterior distribution that allows for a larger step size to be used. This behavior is well known in the context of trajectory optimization: shooting methods often have more difficulty converging than collocation methods due to the highly nonlinear dependence of the objective function on the control variables and initial conditions \cite{Betts_1998}. 

In contrast to ensemble MCMC and MALA, NUTS performs slightly worse than CMALA on the three-species repressilator and slightly better than CMALA on the seven-species repressilator. This can be attributed to the use of preconditioning in NUTS: whereas the constrained and unconstrained Langevin samplers use a unit mass matrix, NUTS uses the empirical covariance computed over a warmup period as the inverse mass matrix. We expect that the performance of the constrained Langevin sampler can be further improved if its dynamics are preconditioned as well.

Interestingly, we obtain a further 15 fold speedup when we do not adjust the constrained Langevin dynamics with the Metropolis criterion (the constrained unadjusted Langevin algorithm, CULA, in Table \ref{tab:performance_metrics}), at the expense of some bias due to the numerical error of discretizing the Langevin equations. We believe that this is due to the reversal of the momentum when steps are rejected in the MALA dynamics. The lower acceptance rate of MALA over CULA causes the MALA dynamics to retrace previously explored regions more often when the noise in the dynamics is small, as is the case for our calculations.

Because the constrained and unconstrained methods treat the initial conditions differently, their parameter distributions should not be exactly the same. Nevertheless, they contain similar information: most of the normalized importance weights (see Section \ref{sec:importance_sampling}) for reweighting the distribution sampled by the constrained Langevin dynamics to the distribution sampled by the unconstrained methods are close to one (and only a tiny fraction are larger than three, as shown in  Figure \ref{fig:hist_emcee_interpolated}, representing the unconstrained methods by ensemble MCMC). Given this similarity, the marginal distribution of \(\mathbf k\) for the constrained Langevin sampler can be reweighted to that for the unconstrained methods or vice versa if desired.

\section{Conclusions}\label{sec:conclusion}

Here, we introduced a constrained Langevin sampler that enables harvesting features of a dynamical system such as fixed points and limit cycles without integrating the dynamics for long times. The sampler is closely related to ones used previously in molecular simulations \cite{Andersen_1983, Barth_1995, Lee_2005, Leimkuhler_2016,Xu_2024}, but to the best of our knowledge, the present study is the first to apply it to parameter estimation.

To converge the parameter distribution for a model of a biochemical oscillator fit to time-series data, our constrained sampler needed orders of magnitude fewer likelihood evaluations than unconstrained samplers without preconditioning and comparable numbers of likelihood evaluations for appropriately preconditioned unconstrained samplers. This suggests the conditioning of posterior distributions associated with ODE parameter estimation problems can be improved by formulating the ODEs as constraints to be satisfied during sampling. In the specific case of oscillator models, the conditioning of posterior distributions is significantly improved by eliminating the dependence of the phase of the model output on model parameters, a known problem when sampling oscillator model parameters \cite{Hwang_2025}. For simplicity, we did not tune the inverse mass matrix for the constrained dynamics; we note that existing methods for estimating the inverse mass matrix based on Euclidean distances \cite{Leimkuhler_2017,Garbuno_Inigo_2020} might not perform well without modification because the constraint manifold is nonlinear. That said, we expect that with appropriate preconditioning the constrained sampler can outperform unconstrained MCMC in general. 

The main challenge in applying our method to new problems is in formulating and solving the constraint equations. For example, different defining systems for bifurcations \cite{Govaerts_2000}, although formally equivalent, may exhibit different numerical behavior. We note that, when incorporating continuous constraints such as those arising from a boundary value problem, the choice of discretization and solution method can greatly affect the efficiency of the sampling procedure. If the mesh is too coarse, the solution to the boundary value problem may fail to converge and result in many rejected steps, whereas, if the mesh is too fine, the constraint equations may be computationally expensive to solve. In our implementation, the number of mesh points is fixed, but their spacing adapts to concentrate them where the solution has the highest curvature. We expect that the performance of the method can be further improved with more sophisticated methods \cite{Carey_1997, Weirs_2005}.

Our focus here is on ODEs, but the method can be extended to systems of partial differential equations (PDEs) by spatially discretizing the system of PDEs and then incorporating the resulting system of ODEs as constraints. As we have done with ODEs, the limiting behaviors of a system of PDEs can be obtained with much less computational effort than numerically integrating in time as is done in \cite{Campillo_Funollet_2018, Kazarnikov_2020}. The choice of spatial discretization, however, may have a large effect on the efficiency of the sampling scheme \cite{Cotter_2013, Bui_Thanh_2014, Ottobre_2016, Bui_Thanh_2016}.

Because the method does not rely on a specific form of the likelihood, it can be combined with enhanced sampling methods \cite{Dinner_2020,Matthews_2018,Neal_2001,Murakami_2014,Pullen_2014,Swendsen_1986} for further speedups.  Recently, machine-learning methods have also been suggested for MCMC for parameter estimation \cite{Gabrie_2022,Wang_2019,Cranmer_2020}. In principle, the logarithm of a learned parameter distribution could be used as an additional potential term to bias the constrained Langevin dynamics. Thorough sampling of parameter distributions of complex models can open the door to systematic assessments of models in biology and other fields where they have been limited to date.

\section{Methods}\label{sec:methods}

\subsection{Numerical integration of Hamilton's equations of motion with constraints}\label{sec:split}
Following \cite{Leimkuhler_2016}, the Hamiltonian can be split into terms corresponding to the potential and kinetic energies:
\begin{subequations}
\begin{align}
    \mathcal H_A &= \frac{1}{2}\mathbf p^{\mathsf T}\mathbf M^{-1}\mathbf p\\
    \mathcal H_B &= U(\mathbf q),
\end{align}\label{eqn:split_hamiltonian}
\end{subequations}
The equations of motion for a system with Hamiltonian \(\mathcal H_A\) are
\begin{equation}
        \begin{aligned}
       \dot{\mathbf q} &= \mathbf M^{-1}\mathbf p\\
       \dot{\mathbf p} &= -\mathbf c_\mathbf q(\mathbf q)^{\mathsf T}\boldsymbol\lambda\\
       \mathbf 0 &= \mathbf c(\mathbf q),
   \end{aligned}\label{eqn:A}
\end{equation}
and the equations of motion for a system with Hamiltonian \(\mathcal H_B\) are
\begin{equation}
    \begin{aligned}
           \dot{\mathbf q} &= \mathbf 0\\
           \dot{\mathbf p} &= -U_{\mathbf q}(\mathbf q)-\mathbf c_\mathbf q(\mathbf q)^{\mathsf T}\boldsymbol\mu\\
           \mathbf 0 &= \mathbf c_{\mathbf q}(\mathbf q)\mathbf M^{-1}\mathbf p,
       \end{aligned}\label{eqn:B}
\end{equation}
where \(\boldsymbol \lambda\) and \(\boldsymbol\mu\) are the Lagrange multipliers appearing in the kinetic and potential terms, respectively. The dynamics for (\ref{eqn:A}) and (\ref{eqn:B}) are discretized separately. Discretizing (\ref{eqn:A}) gives
\begin{subequations}
    \begin{align}
         \mathbf q(t + \Delta t) &= \mathbf q(t) + \Delta t\mathbf M^{-1}{\mathbf p}(t + \Delta t)\label{eqn:A_position_update}\\
         {\mathbf p}(t + \Delta t) &= \mathbf p(t) - \Delta t\mathbf c_{\mathbf q}(\mathbf q(t))^{\mathsf T}\boldsymbol \lambda (t + \Delta t)\label{eqn:A_momentum_update}\\
        \mathbf 0 &= \mathbf c(\mathbf q(t + \Delta t)),
        \label{eqn:A_position_constraint}
    \end{align}\label{eqn:A_discrete}
\end{subequations}
and discretizing (\ref{eqn:B}) gives
\begin{subequations}
    \begin{align}
        \mathbf q(t + \Delta t) &= \mathbf q(t)\\
        \mathbf p(t + \Delta t) &= \mathbf p(t) - \Delta t\left(U_\mathbf q(\mathbf q(t)) + \mathbf c_{\mathbf q}(\mathbf q(t))^{\mathsf T}\boldsymbol\mu(t + \Delta t)\right)\label{eqn:B_momentum_update}\\
        \mathbf 0 &= \mathbf c_{\mathbf q}(\mathbf q(t))\mathbf M^{-1}\mathbf p(t + \Delta t).\label{eqn:B_cotangency_constraint}
    \end{align}\label{eqn:B_discrete}
\end{subequations}
The momentum update \eqref{eqn:B_discrete} is an explicit system of linear equations that can be solved directly, whereas the position update \eqref{eqn:A_discrete} is a nonlinear system of equations that must be solved iteratively. The details of solving these equations are discussed in section \ref{sec:nonlinear_solve}. 

At the start of each step, \(\mathbf q(t)\) is assumed to satisfy the position constraints \(\mathbf c(\mathbf q(t)) = \mathbf 0\) and \(\mathbf p(t)\) is assumed to satisfy the cotangency constraints \(\mathbf c_\mathbf q(\mathbf q(t))\mathbf M^{-1}\mathbf p(t)=0\). For a given timestep \(\Delta t\), (\ref{eqn:A_discrete}) and (\ref{eqn:B_discrete}) define discrete maps \(\Phi_{A, \Delta t}\) and \(\Phi_{B, \Delta t}\) (Algorithms \ref{alg:A} and \ref{alg:B}) that can be composed to approximate the continuous flow map (\ref{eqn:hamilton_eqs}). Repeatedly applying \(\Phi_{B,\Delta t/2}\Phi_{A,\Delta t}\Phi_{B,\Delta t/2}\), for example, results in the WIGGLE integrator~\cite{Lee_2005}.

\begin{algorithm}
    \caption{\texttt{constrained\_position\_step} (\(\Phi_{A, \Delta t}\))\label{alg:A}}
    \textbf{Input: }{position \(\mathbf q_i\), momentum \(\mathbf p_i\), step size \(\Delta t\), inverse mass matrix \(\mathbf M^{-1}\), constraint function \(\mathbf c(\cdot)\)}\\
    \textbf{Output: }{new position \(\mathbf q_{i + 1}\), new momentum \(\mathbf p_{i + 1}\)}
    \begin{algorithmic}[1]
    \State \(\boldsymbol \lambda _{i+1}\leftarrow \texttt {nonlinear\_solve}\left [\mathbf c\left (\mathbf q_i + \Delta t\mathbf M^{-1}\left (\mathbf p_i - \Delta t\mathbf c_{\mathbf q}(\mathbf q_i)^{\mathsf T}\boldsymbol \lambda _{i+1}\right )\right ) = \mathbf 0\right ]\)
    \State \(\mathbf p_{i+1} \leftarrow \mathbf p_i - \Delta t\mathbf c_{\mathbf q}(\mathbf q_i)^{\mathsf T}\boldsymbol \lambda _{i+1}\)
    \State \(\mathbf q_{i + 1}\leftarrow \mathbf q_{i} + \Delta t\mathbf M^{-1}\mathbf p_{i + 1}\)\;
    \State \textbf{return }{\(\mathbf q_{i + 1}, \mathbf p_{i + 1}\)}
    \end{algorithmic}
\end{algorithm}

\begin{algorithm}
    \caption{\texttt{constrained\_momentum\_step} (\(\Phi_{B, \Delta t}\))\label{alg:B}}
    \textbf{Input: }{position \(\mathbf q_i\), momentum \(\mathbf p_i\), step size \(\Delta t\), inverse mass matrix \(\mathbf M^{-1}\), constraint function \(\mathbf c(\cdot)\), potential energy function \(U(\cdot)\)}\\
    \textbf{Output: }{new position \(\mathbf q_{i + 1}\), new momentum \(\mathbf p_{i + 1}\)}
    \begin{algorithmic}[1]
        \State \(\boldsymbol \mu _{i+1} \leftarrow -\left (\mathbf c_{\mathbf q}(\mathbf q_i)\mathbf M^{-1}\mathbf c_{\mathbf q}(\mathbf q_i)^{\mathsf T}\right )^{-1}\mathbf c_{\mathbf q}(\mathbf q_i)\mathbf M^{-1}U_{\mathbf q}(\mathbf q_i) / \Delta t\)
        \State \(\mathbf p_{i+1} \leftarrow \mathbf p_{i} - \Delta t\left (U_{\mathbf q}(\mathbf q_i) + \mathbf c_{\mathbf q}(\mathbf q_i)^{\mathsf T}\boldsymbol \mu _{i+1}\right )\)
        \State \textbf{return }{\(\mathbf q_{i}, \mathbf p_{i + 1}\)}
    \end{algorithmic}
\end{algorithm}


\subsection{Numerical integration of constrained Langevin equations}\label{sec:obabo}

Just as we added noise and friction terms in \eqref{eqn:langevin}, we can add a term corresponding to an Ornstein--Uhlenbeck process in \(\mathbf p\) to (\ref{eqn:A}) and (\ref{eqn:B}) to make the dynamics stochastic:
\begin{equation}
    \begin{aligned}
        \dot{\mathbf q} &= 0\\
        \dot{\mathbf p} &= -\gamma\mathbf p + \sqrt{2T\gamma}\mathbf M^{1/2}\boldsymbol\eta(t) - \mathbf c_{\mathbf q}(\mathbf q)^{\mathsf T}\boldsymbol\mu\\
        \mathbf 0 &= \mathbf c_{\mathbf q}(\mathbf q)\mathbf M^{-1}\mathbf p
    \end{aligned}
    \label{eqn:O}
\end{equation}
These equations can be integrated exactly (in the sense of averages) over a time interval \(\Delta t\):
\begin{subequations}
    \begin{align}
        \mathbf q(t + \Delta t) &= \mathbf q(t)\\
        \mathbf p(t + \Delta t) &= a\mathbf p(t) + b\mathbf M^{1/2}\mathbf r - \Delta t\mathbf c_{\mathbf q}(\mathbf q(t))^{\mathsf T}\boldsymbol \mu (t + \Delta t)\\
        \mathbf 0 &= \mathbf c_{\mathbf q}(\mathbf q(t))\mathbf M^{-1}\mathbf p(t + \Delta t),
    \end{align}\label{eqn:O_discrete}
\end{subequations}
where $a\equiv \exp(-\gamma\Delta t)$, $b\equiv \sqrt{T(1 - a^2)}$, and $\mathbf r\sim \mathcal N(\mathbf 0, \mathbf I)$.
\eqref{eqn:O_discrete} defines a stochastic map \(\Phi_{O,\Delta t}\) (Algorithm \ref{alg:O}). 

The $\Phi_{A,\Delta t}$, $\Phi_{B,\Delta t}$, and $\Phi_{O,\Delta t}$ maps are combined in various ways in \cite{Leimkuhler_2016}. Among them, we choose to use the ``OBABO'' scheme which  approximates (\ref{eqn:langevin}) by the map \(\Phi_{O,\Delta t/2}\Phi_{B,\Delta t/2}\) \(\Phi_{A,\Delta t}\Phi_{B,\Delta t/2}\Phi_{O,\Delta t/2}\) because it minimizes the number of applications of the $\Phi_{A,\Delta t}$ map, which is the most computationally expensive owing to its iterative solution of the constraint equations.  

\begin{algorithm}
    \caption{\texttt{constrained\_momentum\_noise}\label{alg:O} (\(\Phi_{O,\Delta t}\))}
    \textbf{Input: }{position \(\mathbf q_i\), momentum \(\mathbf p_i\), step size \(\Delta t\), inverse mass matrix \(\mathbf M^{-1}\), constraint function \(\mathbf c(\cdot)\), friction coefficient \(\gamma\), temperature \(T\)}\\
    \textbf{Output: }{new position \(\mathbf q_{i + 1}\), new momentum \(\mathbf p_{i + 1}\)}
    \begin{algorithmic}[1]
        \State \(a\leftarrow \exp(-\gamma\Delta t)\)
        \(b\leftarrow \sqrt{T(1 - a^2)}\)\;
        \State \(\mathbf r\sim\mathcal N(0, \mathbf I)\)
        \State \(\boldsymbol \mu _{i+1}\leftarrow b\left (\mathbf c_{\mathbf q}(\mathbf q_i)\mathbf M^{-1}\mathbf c_{\mathbf q}(\mathbf q_i)^{\mathsf T}\right )^{-1}\mathbf c_{\mathbf q}(\mathbf q_i)\mathbf M^{-1/2}\mathbf r / \Delta t\)
        \State \(\mathbf p_{i+1}\leftarrow a\mathbf p_{i} + b\mathbf M^{1/2}\mathbf r - \Delta t\mathbf c_{\mathbf q}(\mathbf q_i)^{\mathsf T}\boldsymbol \mu _{i+1}\)
        \State \textbf{return }{\(\mathbf q_{i}\), \(\mathbf p_{i + 1}\)}
    \end{algorithmic}
\end{algorithm}

The error in the discrete dynamics introduces bias to the sampled distribution. This error can be controlled by reducing the timestep \(\Delta t\). However, it may be preferable to use a large timestep to reduce the amount of computational effort required to generate independent samples. The bias due to discretization errors can be corrected by applying the Metropolis criterion \cite{Metropolis1953}. A particular source of bias that we handle specially is the possibility of non-reversible changes in the position. For large timesteps, the position update \eqref{eqn:A_discrete} may not be reversible up to momentum reversal \cite{Lelievre_2019}; that is, the point obtained by performing a position update and then performing a second position update after reversing the resulting momentum may differ from the initial point. As the convergence of the samplers that we use relies on the reversibility of the dynamics, such errors introduce bias. We correct for non-reversibility by performing a second position update with reversed momentum after every position update and rejecting any steps for which the difference between the initial point and the endpoint of the reverse step is greater than a specified tolerance. One step of the sampling scheme is given by Algorithm \ref{alg:obabo}.

\begin{algorithm}
    \caption{\texttt{obabo\_step}\label{alg:obabo}}
    \textbf{Input:} {position \(\mathbf q_i\), momentum \(\mathbf p_i\), step size \(\Delta t\), inverse mass matrix \(\mathbf M^{-1}\), constraint function \(\mathbf c(\cdot)\), potential energy function \(U(\cdot)\), friction coefficient \(\gamma\), temperature \(T\), reversibility tolerance \(\epsilon_{\textrm{rev}}\)}\\
    \textbf{Output: }{new position \(\mathbf q_{i + 1}\), new momentum \(\mathbf p_{i + 1}\)}
    \begin{algorithmic}[1]
        \State \(\mathbf q_{i + 1}, \mathbf p_{i + 1}\leftarrow \hyperref[alg:O]{\texttt{constrained\_momentum\_noise}}(\mathbf q_i, \mathbf p_i, \Delta t/2,\mathbf M^{-1}, \mathbf c, \gamma, T)\)
        \State \(\mathbf q_{i + 1}, \mathbf p_{i + 1}\leftarrow \hyperref[alg:B]{\texttt{constrained\_momentum\_step}}(\mathbf q_i, \mathbf p_{i + 1}, \Delta t/2, \mathbf M^{-1}, \mathbf c)\)
        \State \(\mathbf q_{i + 1}, \mathbf p_{i + 1}\leftarrow \hyperref[alg:A]{\texttt{constrained\_position\_step}}(\mathbf q_{i}, \mathbf p_{i + 1}, \Delta t, \mathbf M^{-1}, \mathbf c)\)
        \State \(\tilde{\mathbf q}_i, \tilde{\mathbf p}_i, \leftarrow\hyperref[alg:A]{\texttt{constrained\_position\_step}}(\mathbf q_{i + 1}, -\mathbf p_{i + 1}, \Delta t,\mathbf M^{-1}, \mathbf c)\)
        \State \texttt{accept} \(\leftarrow\) \(\left\|\tilde{\mathbf q}_{i} - \mathbf q_{i}\right\| < \epsilon_\textrm{rev}\)\Comment{reversibility check}
        \If{\texttt{accept}}
            \State \(\mathbf q_{i + 1}, \mathbf p_{i + 1}\leftarrow \hyperref[alg:B]{\texttt{constrained\_momentum\_step}}(\mathbf q_{i + 1}, \mathbf p_{i + 1}, \Delta t/2, \mathbf M^{-1}, \mathbf c)\)
            \State \(\mathbf q_{i + 1}, \mathbf p_{i + 1}\leftarrow \hyperref[alg:O]{\texttt{constrained\_momentum\_noise}}(\mathbf q_{i + 1}, \mathbf p_{i + 1}, \Delta t/2,\mathbf M^{-1}, \mathbf c, \gamma, T)\)
            \State \(H_i \leftarrow U(\mathbf q_{i}) + \mathbf p_{i}^{\mathsf T}\mathbf M^{-1}\mathbf p_{i}/2\)\;
            \State \(H_{i + 1} \leftarrow U(\mathbf q_{i + 1}) + \mathbf p_{i + 1}^{\mathsf T}\mathbf M^{-1}\mathbf p_{i + 1}/2\)
            \State \(r \sim U(0,1)\)
            \State \texttt{accept} \(\leftarrow\) \(r < \exp(H_i - H_{i + 1})\)
        \EndIf
        \If{\textbf{not} \texttt{accept}}
            \State \(\mathbf q_{i + 1}, \mathbf p_{i + 1}\leftarrow \mathbf q_{i}, -\mathbf p_{i}\)
        \EndIf
        \State \textbf{return }{\(\mathbf q_{i + 1}, \mathbf p_{i + 1}\)}
    \end{algorithmic}
\end{algorithm}

\subsection{Solving the constraint equations} \label{sec:nonlinear_solve}
\subsubsection{Dense \(\mathbf{c_q}\)}
In computing the momentum update, substituting the expression for \(\mathbf p(t + \Delta t)\) from (\ref{eqn:B_momentum_update}) into (\ref{eqn:B_cotangency_constraint}) gives
\begin{equation}\mathbf c_{\mathbf q}(\mathbf q(t))\mathbf M^{-1}\hat{\mathbf p}(t + \Delta t) - \Delta t\mathbf c_{\mathbf q}(\mathbf q(t))\mathbf M^{-1}\mathbf c_{\mathbf q}(\mathbf q(t))^{\mathsf T}\boldsymbol\mu(t + \Delta t) = \mathbf 0,\label{eqn:linsol_cotangency_1}
\end{equation}
where
\begin{equation}\hat{\mathbf p}(t + \Delta t) = \mathbf p(t) - \Delta t U_{\mathbf q}(\mathbf q(t))\end{equation}
is the momentum after an unconstrained step. Rearranging (\ref{eqn:linsol_cotangency_1}) gives a linear system for \(\boldsymbol \mu(t + \Delta t)\):
\begin{equation}
    \boldsymbol\mu(t + \Delta t) = \frac{1}{\Delta t}(\mathbf c_{\mathbf q}(\mathbf q(t))\mathbf M^{-1}\mathbf c_{\mathbf q}(\mathbf q(t))^\mathsf{T})^{-1}\mathbf c_{\mathbf q}(\mathbf q(t))\mathbf M^{-1}\hat{\mathbf p}(t + \Delta t).\label{eqn:linsol_cotangency_2}
\end{equation}
If \(\mathbf c_{\mathbf q}(\mathbf q(t))\) is dense, an efficient way to solve (\ref{eqn:linsol_cotangency_2}) is to compute the LQ factorization of \(\mathbf c_{\mathbf q}(\mathbf q(t))\mathbf M^{-1/2}\) (i.e., the QR factorization of \(\left(\mathbf M^{-1/2}\right)^{\mathsf{T}}\mathbf c_{\mathbf q}(\mathbf q(t))^\mathsf{T}\)), such that \(\mathbf c_{\mathbf q}(\mathbf q(t))\mathbf M^{-1/2} = \mathbf{LQ}\), where \(\mathbf L\) is a lower triangular matrix, \(\mathbf Q\) is a matrix with orthonormal rows, and \(\mathbf M^{-1/2}\) is a matrix that satisfies \(\mathbf M^{-1/2}\left(\mathbf M^{-1/2}\right)^{\mathsf T} = \mathbf M^{-1}\) (e.g., the Cholesky factor of \(\mathbf M^{-1}\)). Assuming that \(\mathbf c_{\mathbf q}\) has full row rank everywhere, \(\mathbf p(t + \Delta t)\) and \(\boldsymbol\mu(t)\) in terms of \(\mathbf L\) and \(\mathbf Q\) are given by
\begin{subequations}
    \begin{align}
        \boldsymbol\mu(t + \Delta t) &=\frac{1}{\Delta t}(\mathbf c_{\mathbf q}(\mathbf q(t))\mathbf M^{-1}\mathbf c_{\mathbf q}(\mathbf q(t))^\mathsf{T})^{-1}\mathbf c_{\mathbf q}(\mathbf q(t))\mathbf M^{-1}\hat{\mathbf p}(t + \Delta t)\nonumber\\
        &=\frac{1}{\Delta t}(\mathbf c_{\mathbf q}(\mathbf q(t))\mathbf M^{-1}\mathbf c_{\mathbf q}(\mathbf q(t))^\mathsf{T})^{-1}\mathbf c_{\mathbf q}(\mathbf q(t))\mathbf M^{-1/2}\left(\mathbf M^{-1/2}\right)^{\mathsf T}\hat{\mathbf p}(t + \Delta t)\nonumber\\
        &=\frac{1}{\Delta t}\left(\mathbf L\mathbf Q\mathbf Q^{\mathsf{T}}\mathbf L^{\mathsf{T}}\right)^{-1}\mathbf L\mathbf Q\left(\mathbf M^{-1/2}\right)^{\mathsf T}\hat{\mathbf p}(t + \Delta t)\nonumber\\ 
        &=\frac{1}{\Delta t}\mathbf L^{-\mathsf T}\mathbf Q\left(\mathbf M^{-1/2}\right)^{\mathsf T}\hat{\mathbf p}(t + \Delta t)\\
        \mathbf p(t + \Delta t) &= \hat {\mathbf p}(t + \Delta t) -\Delta t\mathbf c_{\mathbf q}(\mathbf q(t))^{\mathsf T}\boldsymbol{\mu }(t + \Delta t) \\ 
        &=\hat {\mathbf p}(t + \Delta t) -\Delta t\left (\mathbf M^{1/2}\right )^{\mathsf T}\left (\mathbf M^{-1/2}\right )^{\mathsf T}\mathbf c_{\mathbf q}(\mathbf q(t))^{\mathsf T}\boldsymbol{\mu }(t + \Delta t)\nonumber \\ 
        &=\hat {\mathbf p}(t + \Delta t) - \left (\mathbf M^{1/2}\right )^{\mathsf T}\mathbf Q^{\mathsf T}\mathbf L^{\mathsf T}\mathbf L^{-\mathsf T}\mathbf Q\left (\mathbf M^{-1/2}\right )^{\mathsf T}\hat {\mathbf p}(t + \Delta t)\nonumber \\ 
        &= \hat {\mathbf p}(t + \Delta t) - \left (\mathbf M^{1/2}\right )^{\mathsf T}\mathbf Q^{\mathsf{T}}\mathbf Q\left (\mathbf M^{-1/2}\right )^{\mathsf T}\hat {\mathbf p}(t + \Delta t). 
    \end{align}\label{eqn:linsol_cotangency_lq}
\end{subequations}

For the position update, substituting (\ref{eqn:A_position_update}) and (\ref{eqn:A_momentum_update}) into (\ref{eqn:A_position_constraint}) gives the nonlinear system of equations in terms of \(\boldsymbol\lambda\):
\begin{equation}
    \mathbf c(\mathbf q(t + \Delta t)) = \mathbf c(\hat {\mathbf q}(t + \Delta t) - \Delta t\mathbf M^{-1}\mathbf c_{\mathbf q}(\mathbf q(t))^{\mathsf{T}}\boldsymbol \lambda (t + \Delta t)) = \mathbf 0,\label{eqn:nlsol_position_1}
\end{equation}
where
\begin{equation}\hat{\mathbf q}(t + \Delta t) = \mathbf q(t) + \Delta t\mathbf M^{-1}\mathbf p(t)\end{equation}
is an unconstrained position update. Differentiating (\ref{eqn:nlsol_position_1}) with respect to \(\boldsymbol\lambda\) gives
\begin{equation}
    \begin{aligned}
        \mathbf c_{\boldsymbol \lambda }(\mathbf q(t + \Delta t)) &= \mathbf c_{\mathbf q}(\mathbf q(t + \Delta t))\mathbf q_{\boldsymbol \lambda }(t + \Delta t)\\ 
        &= -\Delta t\mathbf c_{\mathbf q}(\mathbf q(t + \Delta t))\mathbf M^{-1}\mathbf c_{\mathbf q}(\mathbf q(t))^{\mathsf{T}}. 
    \end{aligned}\label{eqn:jac_constraint}
\end{equation}
\(\boldsymbol\lambda(t + \Delta t)\) and \(\mathbf q(t + \Delta t)\) can then be computed using Newton's method. Given approximations \(\boldsymbol \lambda ^{(i)}(t + \Delta t)\) to \(\boldsymbol \lambda(t + \Delta t)\) and \(\mathbf q^{(i)}(t + \Delta t)\) to \(\mathbf q(t + \Delta t)\), improved approximations can be computed by iteratively solving
\begin{equation}
    \begin{aligned}
         \delta \boldsymbol \lambda ^{(i)} &= \frac {1}{\Delta t}(\mathbf c_{\mathbf q}(\mathbf q^{(i)}(t + \Delta t))\mathbf M^{-1}\mathbf c_{\mathbf q}(\mathbf q(t))^{\mathsf{T}})^{-1}\mathbf c(\mathbf q^{(i)}(t + \Delta t)),\\ 
         \boldsymbol \lambda ^{(i+1)}(t + \Delta t) &= \boldsymbol \lambda ^{(i)}(t + \Delta t) + \delta \boldsymbol \lambda ^{(i)},\\ 
         \mathbf q^{(i+1)}(t + \Delta t) &= \hat {\mathbf q}(t + \Delta t) - \Delta t^2\mathbf M^{-1}\mathbf c_{\mathbf q}(\mathbf q(t))^{\mathsf{T}}\boldsymbol \lambda ^{(i+1)}(t + \Delta t),\\ 
         &\equiv \mathbf q^{(i)}(t + \Delta t) + \delta \mathbf q^{(i)}.
    \end{aligned}\label{eqn:newton_iter_position}
\end{equation}
However, computing and inverting the matrix \(\mathbf c_{\mathbf q}(\mathbf q(t + \Delta t))\mathbf M^{-1}\mathbf c_{\mathbf q}(\mathbf q(t))^\mathsf{T}\) can be quite computationally costly, and depending on the conditioning of \(\mathbf c_{\mathbf q}\), may be numerically unstable. Consequently, we instead approximate \(\mathbf c_{\mathbf q}(\mathbf q(t + \Delta t))\mathbf M^{-1}\mathbf c_{\mathbf q}(\mathbf q(t))^\mathsf{T}\) with \(\mathbf c_{\mathbf q}(\mathbf q(t))\mathbf M^{-1}\mathbf c_{\mathbf q}(\mathbf q(t))^\mathsf{T}\) for small \(\Delta t\) as proposed in \cite{Barth_1995}. As \(\mathbf c_{\mathbf q}(\mathbf q(t))\mathbf M^{-1}\mathbf c_{\mathbf q}(\mathbf q(t))^\mathsf{T}\) is symmetric and assumed to be positive definite, quasi-Newton iterates can be efficiently computed in terms of the LQ factorization of \(\mathbf c_{\mathbf q}(\mathbf q(t))\mathbf M^{-1/2}\):
\begin{subequations}
    \begin{align}
        \delta\boldsymbol\lambda^{(i)} &= \frac{1}{\Delta t}\left(\mathbf c_{\mathbf q}(\mathbf q(t))\mathbf M^{-1}\mathbf c_{\mathbf q}(\mathbf q(t))^\mathsf{T}\right)^{-1}\mathbf c(\mathbf q^{(i)}(t + \Delta t))\nonumber\\
        &=\frac{1}{\Delta t}\left(\mathbf L\mathbf Q\mathbf Q^{\mathsf T}\mathbf L^{\mathsf T}\right)^{-1}\mathbf c(\mathbf q^{(i)}(t + \Delta t))\nonumber\\
        &=\frac{1}{\Delta t}(\mathbf L\mathbf L^{\mathsf T})^{-1}\mathbf c(\mathbf q^{(i)}(t + \Delta t))\\
        \delta \mathbf q^{(i)} &= -\Delta t^2\mathbf M^{-1}\mathbf c_{\mathbf q}(\mathbf q(t))^{\mathsf{T}}\delta \boldsymbol \lambda ^{(i)}\\ 
        &=-\Delta t^2\mathbf M^{-1/2}\left (\mathbf M^{-1/2}\right )^{\mathsf T}\mathbf c_{\mathbf q}(\mathbf q(t))^{\mathsf{T}}\delta \boldsymbol \lambda ^{(i)}\nonumber \\ 
        &=-\Delta t\mathbf M^{-1/2}\mathbf Q^{\mathsf T}\mathbf L^{\mathsf T}(\mathbf L\mathbf L^{\mathsf T})^{-1}\mathbf c(\mathbf q^{(i)}(t + \Delta t))\nonumber \\ 
        &=-\Delta t\mathbf M^{-1/2}\mathbf Q^{\mathsf T}\mathbf L^{-1}\mathbf c(\mathbf q^{(i)}(t + \Delta t)). 
    \end{align}\label{eqn:quasi_newton_iter_position}
\end{subequations}
To improve the rate of convergence, Broyden updates can be applied to the factorization of \(\mathbf c_{\mathbf q}(\mathbf q(t))\mathbf M^{-1}\mathbf c_{\mathbf q}(\mathbf q(t))^\mathsf{T}\). Our implementation computes Broyden updates using the QNERR algorithm described in \cite{Deuflhard_2011}.
\subsubsection{Sparse \(\mathbf c_{\mathbf q}\)}
If \(\mathbf c\) contains constraints arising from a boundary value problem discretized using a local basis, then \(\mathbf c_{\mathbf q}\) can be partitioned as
\begin{equation}\mathbf c_{\mathbf q} = \left(\begin{array}{cc}
    \mathbf c_{\mathbf y}&\mathbf c_{\mathbf k}
\end{array}\right),\end{equation}
where \(\mathbf c_{\mathbf y}\) is a sparse matrix containing derivatives with respect to the variables of the discretized boundary value problem, and \(\mathbf c_{\mathbf k}\) is a dense matrix containing derivatives with respect to parameters. Efficient solution of (\ref{eqn:B_discrete}) would ideally exploit the sparsity of \(\mathbf c_{\mathbf q}\mathbf M^{-1/2}\), where we assume that \(\mathbf M^{-1}\) is such that \(\mathbf c_{\mathbf q}\mathbf M^{-1/2}\) has the same structure as \(\mathbf c_{\mathbf q}\) (e.g., diagonal \(\mathbf M^{-1}\)), so that
\begin{equation}\mathbf c_{\mathbf q}\mathbf M^{-1/2} = \left(\begin{array}{cc}
    \mathbf A&\mathbf B
\end{array}\right)\end{equation}
with \(\mathbf A\) sparse and \(\mathbf B\) dense. 
In general, direct LQ factorization of \(\mathbf c_{\mathbf q}\mathbf M^{-1/2}\) results in a dense \(\mathbf Q\), and since \(\mathbf c_{\mathbf q}\) contains dense columns \(\mathbf c_{\mathbf k}\), \(\mathbf L\) is also dense, making the dense solution procedures (\ref{eqn:linsol_cotangency_lq}) and (\ref{eqn:quasi_newton_iter_position}) expensive for fine discretizations of the boundary value problem. However, if the LQ factorization of \(\mathbf c_{\mathbf y}\) can be computed efficiently, Algorithms 3 and 4 in \cite{George_1984} can be applied to update a sparse factorization of \(\mathbf c_{\mathbf y}\), allowing for efficient solution of the linear systems that arise in the position and momentum update steps.

If exact Newton iterations are used to compute the position update, the matrices
\begin{multline}
    \mathbf c_{\mathbf q}(\mathbf q^{(i)}(t + \Delta t))\mathbf M^{-1}\mathbf c_{\mathbf q}(\mathbf q(t))^\mathsf{T} = \mathbf A(\mathbf q^{(i)}(t + \Delta t))\mathbf A(\mathbf q(t))^\mathsf{T} + \mathbf B(\mathbf q^{(i)}(t + \Delta t))\mathbf B(\mathbf q(t))^\mathsf{T}
\end{multline}
must be inverted. Since \(\mathbf c_{\mathbf k}\) generally has many fewer columns than \(\mathbf c_{\mathbf y}\), the Woodbury matrix identity can in principle be applied, treating \(\mathbf B(\mathbf q(t + \Delta t))\mathbf B(\mathbf q(t))^\mathsf{T}\) as a low rank update to \(\mathbf A(\mathbf q(t + \Delta t))\mathbf A(\mathbf q(t))^\mathsf{T}\). However, \(\mathbf A(\mathbf q(t + \Delta t))\mathbf A(\mathbf q(t))^\mathsf{T}\) tends to be ill-conditioned, which leads to poor convergence of the Newton iterations due to inaccurate solution of the linear system. 
Alternatively, \(\mathbf c_{\mathbf q}(\mathbf q(t))\mathbf M^{-1}\mathbf c_{\mathbf q}(\mathbf q(t))^\mathsf{T}\) can be used as a symmetric approximation to \(\mathbf c_{\mathbf q}(\mathbf q(t + \Delta t))\mathbf M^{-1}\mathbf c_{\mathbf q}(\mathbf q(t))^\mathsf{T}\) for quasi-Newton solution of the position constraints as in the dense case. We use Algorithm 3 from \cite{George_1984} to invert \(\mathbf c_{\mathbf q}(\mathbf q(t))\mathbf M^{-1}\mathbf c_{\mathbf q}(\mathbf q(t))^\mathsf{T}\) and use quasi-Newton iterations with Broyden updates to compute position updates as in the dense case. A similar procedure for solving large, sparse constraint equations is proposed in \cite{Xu_2024} where the Cholesky factorization of \(\mathbf c_{\mathbf q}(\mathbf q(t))\mathbf M^{-1}\mathbf c_{\mathbf q}(\mathbf q(t))^{\mathsf{T}}\) is computed instead of the LQ factorization of \(\mathbf c_{\mathbf q}\mathbf M^{-1/2}\).

\subsection{Discretizing periodic orbits}\label{sec:collocation}
To compute solutions to (\ref{eqn:periodic_bvp}) numerically, a suitable discretization must be applied. As only periodic solutions are of interest, a natural choice of discretization is to apply Galerkin's method with Fourier basis functions (i.e., Fourier collocation). This  can be very efficient if the system of ODEs has smooth solutions. However, if the solutions to the system of ODEs or its derivatives undergo rapid variation, the number of Fourier basis functions required for an accurate approximation can be prohibitively large. Consequently, we instead discretize the system of ODEs using Gauss-Legendre collocation on piecewise polynomials \cite{Krauskopf_2007}. Given a mesh 
\begin{equation}
    \{0 = s_0 < s_1 < \dots < s_N  = \tau\}
\end{equation}
with interval widths 

    \begin{align}
        \Delta s_i &= s_i - s_{i - 1},\qquad 1\leq i\leq N
    \end{align} 

we approximate the solution to \eqref{eqn:periodic_bvp} on the space of piecewise polynomials
\begin{equation}
    \mathcal P_{m,\Delta s} = \left\{\mathbf u\in C[0,\tau]\right.\left|\mathbf u|_{[s_{i -1}, s_i]}\in \mathcal P_{m}\right\}.
\end{equation}
That is, \(\mathcal P_{m,\Delta s}\) consists of continuous functions \(\mathbf u\) on the interval \([0, \tau]\) such that on each mesh subinterval \([s_{i - 1}, s_i]\), \(\mathbf u\) is a polynomial of order up to \(m\). We demand that on each subinterval, \(\mathbf u\) satisfies \eqref{eqn:repressilator_log} at collocation points \(\{z_{i,j}\}\)
\begin{equation}
    \mathbf u'(z_{i, j}) = \mathbf f(\mathbf u(z_{i, j}), \mathbf k),\quad j\in\{0,\dots,m\},\quad i\in \{0,\dots,N - 1\}
\end{equation}
where \(\mathbf u'(s)\) denotes the derivative of \(\mathbf u\) with respect to \(s\). We choose the collocation points to be the scaled and shifted roots of the \(m\)-th Legendre polynomial. Our implementation constructs \(\mathbf u\) with Newton polynomials on each mesh subinterval.

Accurate solution of boundary value problems using a mesh-based discretization requires the mesh to be sufficiently fine at regions where the solution undergoes rapid variation. If the boundary value problem is discretized on a uniform mesh, a large number of mesh elements may be placed in relatively smooth regions of the solution that only require a few points to resolve. To save computational effort, adaptive mesh refinement procedures can be used to generate meshes that concentrate mesh points at rapidly varying regions of the solution without using a fine mesh globally. For applications such as finite element calculations, mesh refinement is typically performed by dividing mesh elements into smaller pieces in regions where an error estimate is high ($h$-adaptivity), which changes the total number of mesh points. Changing the size of the mesh, however, complicates reversibility guarantees. As an alternative, we adapt the mesh by redistributing the mesh elements after each step of the sampler while keeping the total number of mesh elements constant ($r$-adaptivity, moving mesh methods). Central to moving mesh methods is the equidistribution principle: for some strictly positive mesh density function \(\rho(s)\) that is proportional to an error estimate of the solution and mesh points \(\{s_0,\dots,s_{N}\}\), the integral of the mesh density function should be equal on each mesh element of the optimal mesh 
\begin{equation}
    \int_{s_i}^{s_{i + 1}}\rho(s)ds = \frac{1}{N}\int_{s_0}^{s_N}\rho(s)ds = \frac{\sigma}{N}.\label{eqn:equidistributing_mesh}
\end{equation}
The mesh density function we use is based on the curvature of \(\mathbf u(t)\) \cite{Huang_2011}:
\begin{align}
    \rho(s) &= \left(1 + \|\mathbf u''(s)\|^2\right)^{1 / 4}
\end{align}
where \(\mathbf u''(s)\) is the second derivative of \(\mathbf u\) with respect to \(s\).
We choose to solve the mesh equations (\ref{eqn:equidistributing_mesh}) and the boundary value problem (\ref{eqn:periodic_bvp}) simultaneously, leading to the system of equations
\begin{equation}
    \begin{aligned}
    \mathbf u'(z_{i, j}) &= \mathbf f(\mathbf u(z_{i, j}), \mathbf k),&i\in \{0,\dots,N - 1\}j\in\{0,\dots,m\}\\
    \mathbf u(0) &= \mathbf u(\tau )\\
        \int_{s_i}^{s_{i + 1}}\rho(s)ds &= \frac{\sigma}{N},&i\in\{0,\dots,N-1\}
    \end{aligned}
\end{equation}
which we solve using sparse Newton iteration.

\subsection{Unconstrained Monte Carlo methods}\label{sec:unconstrained_mcmc}

In this section, we briefly describe the unconstrained MCMC methods with which we compare our method and refer readers to their original references for more complete discussions.

\subsubsection{Metropolis-adjusted Langevin}
One step of the algorithm we refer to as unconstrained Metropolis-adjusted Langevin consists of one timestep of discretized underdamped Langevin (i.e., second-order Langevin) dynamics followed by application of the Metropolis criterion with negation of the momenta upon rejection. As in \eqref{eqn:split_hamiltonian}, the Hamiltonian \(\mathcal H\) can be split into potential and kinetic parts for which the dynamics can be solved exactly. For an unconstrained system, the equations of motion for the potential and kinetic parts, respectively, are
\begin{equation}
        \begin{aligned}
       \dot{\mathbf q} &= \mathbf M^{-1}\mathbf p\\
       \dot{\mathbf p} &= \mathbf 0,
   \end{aligned}\label{eqn:A_unconstrained}
\end{equation}
and
\begin{equation}
    \begin{aligned}
           \dot{\mathbf q} &= \mathbf 0\\
           \dot{\mathbf p} &= -U_{\mathbf q}(\mathbf q).
       \end{aligned}\label{eqn:B_unconstrained}
\end{equation}
Stochasticity can be introduced as an unconstrained Ornstein--Uhlenbeck process in the momenta 
\begin{equation}
    \begin{aligned}
        \dot{\mathbf q} &= 0\\
        \dot{\mathbf p} &= -\gamma\mathbf p + \sqrt{2T\gamma}\mathbf M^{1/2}\boldsymbol\eta(t),
    \end{aligned}
    \label{eqn:O_unconstrained}
\end{equation}
where \(\gamma\) is a friction coefficient. Equations \cref{eqn:A_unconstrained,eqn:B_unconstrained,eqn:O_unconstrained} define unconstrained maps \(\Phi_{A,\Delta t}\), \(\Phi_{B,\Delta t}\), and \(\Phi_{O,\Delta t}\) that can be composed to numerically integrate the Langevin equations of motion. We use the composition \(\Phi_{B,\Delta t/2}\Phi_{A,\Delta t/2}\Phi_{O,\Delta t}\Phi_{A,\Delta t/2}\Phi_{B,\Delta t/2}\) for the unconstrained Metropolis-adjusted Langevin sampler since it has been observed to produce good statistics for position variables \cite{Leimkuhler_2015}. Note that this method differs from the conventional Metropolis-adjusted Langevin algorithm by the inclusion of momentum variables. This method can be interpreted as generalized Hamiltonian Monte Carlo \cite{Horowitz_1991} with the trajectory length set to a single timestep.

\subsubsection{The no-U-turn sampler}

From a starting point \(\mathbf q\), Hamiltonian Monte Carlo in its simplest form consists of sampling a normally distributed momentum vector \(\mathbf p\), integrating the equations of motion for a fixed trajectory length \(L\), and then accepting or rejecting the last point of the trajectory based on the Metropolis criterion. Since a single trajectory is approximately contained within a particular level set of the energy, certain choices of \(L\) can result in slow sampling due to trajectories forming approximately closed orbits. Let \((\mathbf q_{-}, \mathbf p_{-})\) and \(\mathbf (\mathbf q_{+}, \mathbf p_{+})\) denote the position and momentum at the two endpoints of a trajectory. The no-U-turn sampler \cite{NUTS} adaptively sets the trajectory length by integrating trajectories until the criterion

    \begin{equation}
        \begin{aligned}
            \mathbf p_{+}^\mathsf{T}(\mathbf q_{+} - \mathbf q_{-}) & < 0\\
            \mathbf p_{-}^\mathsf{T}(\mathbf q_{-} - \mathbf q_{+}) & < 0
        \end{aligned}\label{eqn:uturn}
    \end{equation}

is satisfied. When \eqref{eqn:uturn} is true, the momenta at the endpoints projected onto the line joining the endpoints of the trajectory point towards each other, and further integration is likely to decrease the distance between the endpoints of the trajectory. To preserve time reversibility, the implementation of NUTS \cite{cabezas2024blackjax} that we use iteratively doubles the length of the trajectory until \eqref{eqn:uturn} is satisfied or a maximum trajectory length is exceeded and then returns a point in the trajectory with probability proportional to its energy.

In addition to adaptively tuning \(L\), the step size \(\Delta t\) is tuned using dual averaging \cite{Nesterov_2012} to achieve a specified acceptance rate, and the inverse mass matrix is approximated as the empirical covariance matrix computed over a warmup period \cite{stan2024}.

\subsubsection{Ensemble MCMC}
The affine-invariant ensemble sampler that we use efficiently samples poorly scaled distributions by evolving an ensemble of parameter sets (random walkers) and using the spread in the ensemble to guide the directions of moves \cite{Goodman_2010, Foreman_Mackey_2013}.  Specifically, at each step, two walkers $\mathbf x_j$ and $\mathbf x_k$ are selected randomly,  $\mathbf x_j$ is replaced with
\begin{equation}
    \mathbf x^{(0)}_j = \mathbf x_k + Z[\mathbf x_j - \mathbf x_k],\label{eqn:stretch_move}
\end{equation}
and the move is accepted with probability 
\begin{equation}
     \min[1,Z^{N-1}\pi(\mathbf x^{(0)}_j)/\pi(\mathbf x_j)],
\end{equation} where $N$ is the dimension of the parameter space, $Z$ is a random variable with density
\begin{equation}
    p(z)\propto\begin{cases}
        z^{-1/2}&\textrm{if } z\in\left[a^{-1},a\right]\\
        0&\textrm{otherwise,}
    \end{cases}
\end{equation}
and \(a\) is an adjustable step size parameter.  We use \(a=1.5\), which we found to give relatively good behavior in preliminary simulations.

\subsection{Accounting for manifold curvature}
\label{sec:reweight_curvature}
The solution manifold to a system of constraints is generally curved, and projecting points sampled on the constraint manifold onto a limited set of variables concentrates the points where the manifold is close to orthogonal to the coordinate axes. For example, projecting equispaced points on the unit circle onto the interval \([-1, 1]\) results in the Chebyshev nodes, which are concentrated near the endpoints \(-1\) and \(1\), where the unit circle becomes increasingly orthogonal to the real line. If this concentration of points is undesired, one can correct for it by weighting each point by a factor that accounts for the change of the volume element of the manifold upon projection.

Given an \(m\)-dimensional submanifold \(\mathcal M\) of \(\mathbb R^{n}\) and a parameterization \(\boldsymbol{\phi}:\mathbb R^{m}\rightarrow\mathcal M\), the volume element of a point \(\boldsymbol{\phi}(\mathbf x)\) on \(\mathcal M\) is given by \(\mathop{\mathrm{det}}(\boldsymbol{\phi}_{\mathbf x}^\mathsf{T}\boldsymbol{\phi}_{\mathbf x})^{1/2}d\mathbf{x}\), where \(\boldsymbol{\phi}_{\mathbf x}\) denotes the matrix of derivatives with elements \(\partial \boldsymbol{\phi}_i/\partial x_j\). The columns of \(\boldsymbol{\phi}_{\mathbf x}(\mathbf x)\) are a basis for the tangent space of \(\mathcal M\) at \(\mathbf x\). We compute the correction for the effect of manifold curvature by constructing an orthonormal basis \(\mathbf Q\in\mathbb R^{n\times m}\) for the tangent space of \(\mathcal M\) at each sampled point and then projecting each column of \(\mathbf Q\) onto \(k\) coordinate axes of interest to obtain \(\hat{\mathbf Q}\in\mathbb R^{k\times m}\), where \(k < m\). The factor by which volume is changed by the projection is given by \(\mathop{\mathrm{det}}(\hat{\mathbf Q}\hat{\mathbf Q}^{\mathsf T})^{1/2}\), which can be used to weight points when constructing a histogram.

\subsection{Assessing convergence}\label{sec:GelmanRubin}

We assess the convergence of simulations using a version of the potential scale reduction factor ($\hat{R}$) defined in \cite{Gelman_1992}.
For a set of \(M\) independent chains each with \(N\) samples, we first compute the within-chain mean ($\bar{\mathbf k}_m$) and covariance ($\boldsymbol\Sigma_{a}$):
\begin{subequations}
\begin{align}
        \bar{\mathbf k}_m &= \frac{1}{N}\sum_{n=1}^N\mathbf k_{mn}\\
        \boldsymbol\Sigma_{a} &= \frac{1}{M}\sum_{m=1}^M \boldsymbol\Sigma_{m}\\
        \boldsymbol\Sigma_{m} &= \frac{1}{N-1}\sum_{n=1}^N\left(\mathbf k_{mn} - \bar{\mathbf k}_{m}\right)\left(\mathbf k_{mn} - \bar{\mathbf k}_{m}\right)^{\mathsf T}.
        \end{align}
\end{subequations}
Then we compute the between-chain covariance ($\boldsymbol\Sigma_{b}$):
\begin{subequations}
\begin{align}
        \boldsymbol\Sigma_{b} &= \frac{N}{M-1}\sum_{m=1}^{M}\left(\bar{\mathbf k}_m - \bar{\mathbf k}\right)\left(\bar{\mathbf k}_m - \bar{\mathbf k}\right)^\mathsf{T}\\
        \bar{\mathbf k} &= \frac{1}{M}\sum_{m=1}^M\bar{\mathbf k}_m.
\end{align}
\end{subequations}
Given the within-chain and between-chain covariances, we estimate the stationary covariance as
\begin{equation}
        \boldsymbol\Sigma = \frac{N - 1}{N}\boldsymbol\Sigma_{a} + \frac{1}{N}\boldsymbol\Sigma_{b}.
\end{equation}
Then,
\begin{equation}
        \hat{R}= \|\boldsymbol\Sigma_{a}^{-1}\boldsymbol\Sigma\|_2.
\end{equation}
The essential idea is that upon convergence, the initial conditions should be forgotten, and the within-chain covariance should equal the stationary covariance; in this case, $\hat{R} = 1$.

\subsection{Importance sampling}\label{sec:importance_sampling}

Given two absolutely continuous probability distributions \(\pi_1\) and \(\pi_2\) defined over the same sample space, the importance weight
\begin{equation}
    w(x) = \frac{\pi_2(x)}{\pi_1(x)}\label{eqn:importance_weight}
\end{equation}
allows for integrals with respect to \(\pi_2\) to be computed using \(\pi_1\). For any function \(f(x)\),
\begin{equation}
    \begin{aligned}
        \mathbb E_{\pi_2}(f(x))\equiv\int f(x)\pi_2(x)\mathop{dx} =\int f(x)\pi_1(x)w(x)\mathop{dx}
        \equiv \mathbb E_{\pi_1}(f(x)w(x)),
    \end{aligned}\label{eqn:reweight}
\end{equation}
where \(w(x)\) is the Radon--Nikodym derivative of \(\pi_2\) with respect to \(\pi_1\). Given samples \(\{x_1,\dots,x_N\}\) from \(\pi_1\), an unbiased estimate of \eqref{eqn:reweight} is
\begin{equation}
    \mathbb E_{\pi_2}(f(x)) \approx \frac{1}{N}\sum_{i=1}^Nf(x_i)\frac{\pi_2(x_i)}{\pi_1(x_i)}
\end{equation}
If \(\pi_1\) and \(\pi_2\) are only known up to a normalization constant, such that
\begin{equation}
    \begin{aligned}
        \pi_1(x) &= c_1\hat{\pi}_1(x)\\
        \pi_2(x) &= c_2\hat{\pi}_2(x),
    \end{aligned}
\end{equation}
then the expectation with respect to \(\pi_2\) can instead be computed as
\begin{equation}
    \begin{aligned}
        \mathbb E_{\pi_2}(x)&= \int f(x) \pi_1(x)\frac{\pi_2(x)}{\pi_1(x)}\mathop{dx}\\
        &=\left.\int f(x) \pi_1(x)\frac{\pi_2(x)}{\pi_1(x)}\mathop{dx}\right/\int \pi_1(x)\frac{\pi_2(x)}{\pi_1(x)}\mathop{dx}\\
        &=\left.\int f(x) \pi_1(x)\frac{\hat{\pi}_2(x)}{\hat{\pi}_1(x)}\mathop{dx}\right/\int \pi_1(x)\frac{\hat{\pi}_2(x)}{\hat{\pi}_1(x)}\mathop{dx}\\
        &\equiv \frac{\mathbb E_{\pi_1}(f(x)\hat{w}(x))}{\mathbb E_{\pi_1}(\hat{w}(x))}.
    \end{aligned}\label{eqn:reweight_normalized}
\end{equation}
Accordingly, we refer to
\begin{equation}
    \frac{\hat{w}(x)}{\mathbb E_{\pi_1}(\hat{w}(x))}
\end{equation}
as the normalized importance weights (cf.\ Figure     \ref{fig:hist_emcee_interpolated}).
Given samples \(\{x_1,\dots,x_N\}\) from \(\pi_1\), a biased estimate for \eqref{eqn:reweight_normalized} is
\begin{equation}
    \frac{\mathbb E_{\pi_1}(f(x)\hat{w}(x))}{\mathbb E_{\pi_1}(\hat{w}(x))} = \left(\sum_{i=1}^Nf(x_i)\hat{w}(x_i)\right)\left(\sum_{i=1}^N\hat{w}(x_i)\right)^{-1}.
\end{equation}

\subsection*{Software}

Open-source software implementing the method using the Google JAX package \cite{jax2018github} is available at \url{https://github.com/dinner-group/constrained-Langevin}.

\section*{Acknowledgments}

We thank Michael Rust, Lu Hong, Yujia Liu, Yihang Wang, Monika Scholz, Elizabeth Jerison, and Nicolas Romeo for helpful discussions.  The authors acknowledge the University of Chicago’s Research Computing Center for computing resources.

\section*{Funding}
A.R.D. acknowledges support from the National Science Foundation (grant number MCB-1953402) and from the National Institute for Theory and Mathematics in Biology, which is funded by the National Science
Foundation (grant number DMS-2235451) and the Simons Foundation (grant number MPTMPS-00005320).  J.Q.W. and A.R.D. acknowledge support from the National Science Foundation (grant number DMS-2054306).


\bibliographystyle{siamplain}
\bibliography{bibliography}

\begin{thebibliography}{10}

\bibitem{Andersen_1983}
{\sc H.~C. Andersen}, {\em Rattle: A
  {\textquotedblleft}velocity{\textquotedblright} version of the shake
  algorithm for molecular dynamics calculations}, Journal of Computational
  Physics, 52 (1983), pp.~24--34.

\bibitem{Barth_1995}
{\sc E.~Barth, K.~Kuczera, B.~Leimkuhler, and R.~D. Skeel}, {\em Algorithms for
  constrained molecular dynamics}, Journal of Computational Chemistry, 16
  (1995), pp.~1192--1209.

\bibitem{Battogtokh_2002}
{\sc D.~Battogtokh, D.~K. Asch, M.~E. Case, J.~Arnold, and H.-B.
  Sch{\"u}ttler}, {\em An ensemble method for identifying regulatory circuits
  with special reference to the qa gene cluster of {Neurospora} crassa},
  Proceedings of the National Academy of Sciences, 99 (2002), pp.~16904--16909.

\bibitem{Beskos_2013}
{\sc A.~Beskos, N.~Pillai, G.~Roberts, J.-M. Sanz-Serna, and A.~Stuart}, {\em
  Optimal tuning of the hybrid {Monte} {Carlo} algorithm}, Bernoulli, 19
  (2013), p.~1501–1534.

\bibitem{Betts_1998}
{\sc J.~T. Betts}, {\em Survey of numerical methods for trajectory
  optimization}, Journal of Guidance, Control, and Dynamics, 21 (1998),
  p.~193–207.

\bibitem{jax2018github}
{\sc J.~Bradbury, R.~Frostig, P.~Hawkins, M.~J. Johnson, C.~Leary,
  D.~Maclaurin, G.~Necula, A.~Paszke, J.~Vander{P}las, S.~Wanderman-{M}ilne,
  and Q.~Zhang}, {\em {JAX}: composable transformations of {P}ython+{N}um{P}y
  programs}, 2018, \url{http://github.com/jax-ml/jax}.

\bibitem{Brown_2003}
{\sc K.~S. Brown and J.~P. Sethna}, {\em Statistical mechanical approaches to
  models with many poorly known parameters}, Physical Review E, 68 (2003),
  p.~021904.

\bibitem{Brubaker_2012}
{\sc M.~Brubaker, M.~Salzmann, and R.~Urtasun}, {\em A family of {M}{C}{M}{C}
  methods on implicitly defined manifolds}, Proceedings of the Fifteenth
  International Conference on Artificial Intelligence and Statistics, PMLR, 22
  (2012), pp.~161--172.

\bibitem{Bui_Thanh_2014}
{\sc T.~Bui-Thanh and M.~Girolami}, {\em Solving large-scale
  {P}{D}{E}-constrained {Bayesian} inverse problems with {Riemann} manifold
  {Hamiltonian} {Monte} {Carlo}}, Inverse Problems, 30 (2014), p.~114014.

\bibitem{Bui_Thanh_2016}
{\sc T.~Bui-Thanh and Q.~P. Nguyen}, {\em {F}{E}{M}-based
  discretization-invariant {M}{C}{M}{C} methods for {P}{D}{E}-constrained
  {Bayesian} inverse problems}, Inverse Problems and Imaging, 10 (2016),
  p.~943–975.

\bibitem{cabezas2024blackjax}
{\sc A.~Cabezas, A.~Corenflos, J.~Lao, R.~Louf, A.~Carnec, K.~Chaudhari,
  R.~Cohn-Gordon, J.~Coullon, W.~Deng, S.~Duffield, G.~Durán-Martín,
  M.~Elantkowski, D.~Foreman-Mackey, M.~Gregori, C.~Iguaran, R.~Kumar, M.~Lysy,
  K.~Murphy, J.~C. Orduz, K.~Patel, X.~Wang, and R.~Zinkov}, {\em
  Black{J}{A}{X}: Composable {B}ayesian inference in {JAX}}, arXiv preprint
  arXiv:2402.10797,  (2024).

\bibitem{Campillo_Funollet_2018}
{\sc E.~Campillo-Funollet, C.~Venkataraman, and A.~Madzvamuse}, {\em {Bayesian}
  parameter identification for {Turing} systems on stationary and evolving
  domains}, Bulletin of Mathematical Biology, 81 (2018), p.~81–104.

\bibitem{Carey_1997}
{\sc G.~F. Carey}, {\em Computational Grids: Generation, Adaptation, and
  Solution Strategies}, CRC Press, Boca Raton, United States, 1997.

\bibitem{Keller_1982}
{\sc T.~F.~C. Chan and H.~B. Keller}, {\em Arc-length continuation and
  multigrid techniques for nonlinear elliptic eigenvalue problems}, SIAM
  Journal on Scientific and Statistical Computing, 3 (1982), pp.~173--194.

\bibitem{Cotter_2013}
{\sc S.~L. Cotter, G.~O. Roberts, A.~M. Stuart, and D.~White}, {\em {MCMC}
  methods for functions: Modifying old algorithms to make them faster},
  Statistical Science, 28 (2013).

\bibitem{Cranmer_2020}
{\sc K.~Cranmer, J.~Brehmer, and G.~Louppe}, {\em The frontier of
  simulation-based inference}, Proceedings of the National Academy of Sciences,
  117 (2020), pp.~30055--30062.

\bibitem{Deuflhard_2011}
{\sc P.~Deuflhard}, {\em Newton Methods for Nonlinear Problems}, Springer
  Berlin Heidelberg, Berlin, Germany, 2011.

\bibitem{Dinner_2020}
{\sc A.~R. Dinner, E.~H. Thiede, B.~V. Koten, and J.~Weare}, {\em
  Stratification as a general variance reduction method for {Markov} chain
  {Monte} {Carlo}}, SIAM/ASA Journal on Uncertainty Quantification, 8 (2020),
  pp.~1139--1188.

\bibitem{Elowitz-2000}
{\sc M.~B. Elowitz and S.~Leibler}, {\em A synthetic oscillatory network of
  transcriptional regulators}, Nature, 403 (2000), pp.~335--338.

\bibitem{Eydgahi_2013}
{\sc H.~Eydgahi, W.~W. Chen, J.~L. Muhlich, D.~Vitkup, J.~N. Tsitsiklis, and
  P.~K. Sorger}, {\em Properties of cell death models calibrated and compared
  using {Bayesian} approaches}, Molecular Systems Biology, 9 (2013), p.~644.

\bibitem{Flaherty_2008}
{\sc P.~Flaherty, M.~L. Radhakrishnan, T.~Dinh, R.~A. Rebres, T.~I. Roach,
  M.~I. Jordan, and A.~P. Arkin}, {\em A dual receptor crosstalk model of
  {G}-protein-coupled signal transduction}, PLOS Computational Biology, 4
  (2008), p.~e1000185.

\bibitem{Foreman_Mackey_2013}
{\sc D.~Foreman-Mackey, D.~W. Hogg, D.~Lang, and J.~Goodman}, {\em emcee: The
  {MCMC} hammer}, Publications of the Astronomical Society of the Pacific, 125
  (2013), pp.~306--312.

\bibitem{Gabrie_2022}
{\sc M.~Gabri{\'e}, G.~M. Rotskoff, and E.~Vanden-Eijnden}, {\em Adaptive
  {Monte} {Carlo} augmented with normalizing flows}, Proceedings of the
  National Academy of Sciences, 119 (2022), p.~e2109420119.

\bibitem{Garbuno_Inigo_2020}
{\sc A.~Garbuno-Inigo, N.~Nüsken, and S.~Reich}, {\em Affine invariant
  interacting {Langevin} dynamics for {Bayesian} inference}, SIAM Journal on
  Applied Dynamical Systems, 19 (2020), p.~1633–1658.

\bibitem{Gelman_1992}
{\sc A.~Gelman and D.~B. Rubin}, {\em Inference from iterative simulation using
  multiple sequences}, Statistical Science, 7 (1992).

\bibitem{George_1984}
{\sc A.~George, M.~T. Heath, and E.~Ng}, {\em Solution of sparse
  underdetermined systems of linear equations}, {SIAM} Journal on Scientific
  and Statistical Computing, 5 (1984), pp.~988--997.

\bibitem{Geyer_1992}
{\sc C.~J. Geyer}, {\em Practical {Markov} chain {Monte} {Carlo}}, Statistical
  Science, 7 (1992).

\bibitem{Girolami_2011}
{\sc M.~Girolami and B.~Calderhead}, {\em Riemann manifold {Langevin} and
  {Hamiltonian} {Monte} {Carlo} methods}, Journal of the Royal Statistical
  Society Series B: Statistical Methodology, 73 (2011), p.~123–214.

\bibitem{Goodman_2010}
{\sc J.~Goodman and J.~Weare}, {\em Ensemble samplers with affine invariance},
  Communications in Applied Mathematics and Computational Science, 5 (2010),
  pp.~65--80.

\bibitem{Govaerts_2000}
{\sc W.~J.~F. Govaerts}, {\em Numerical Methods for Bifurcations of Dynamical
  Equilibria}, Society for Industrial and Applied Mathematics, Philadelphia,
  United States, 2000.

\bibitem{Gutenkunst_2007}
{\sc R.~N. Gutenkunst, J.~J. Waterfall, F.~P. Casey, K.~S. Brown, C.~R. Myers,
  and J.~P. Sethna}, {\em Universally sloppy parameter sensitivities in systems
  biology models}, PLOS Computational Biology, 3 (2007), p.~e189.

\bibitem{Hoffman2021AnAS}
{\sc M.~D. Hoffman, A.~Radul, and P.~Sountsov}, {\em An adaptive-{M}{C}{M}{C}
  scheme for setting trajectory lengths in {{Hamiltonian}} {Monte} {Carlo}}, in
  International Conference on Artificial Intelligence and Statistics, 2021.

\bibitem{NUTS}
{\sc M.~D. Homan and A.~Gelman}, {\em The {N}o-{U}-{T}urn sampler: Adaptively
  setting path lengths in {Hamiltonian} {Monte} {Carlo}}, Journal of Machine
  Learning Research, 15 (2014), p.~1593–1623.

\bibitem{Hong_2020}
{\sc L.~Hong, D.~O. Lavrentovich, A.~Chavan, E.~Leypunskiy, E.~Li, C.~Matthews,
  A.~LiWang, M.~J. Rust, and A.~R. Dinner}, {\em {Bayesian} modeling reveals
  metabolite-dependent ultrasensitivity in the cyanobacterial circadian clock},
  Molecular Systems Biology, 16 (2020).

\bibitem{Horowitz_1991}
{\sc A.~M. Horowitz}, {\em A generalized guided {Monte} {Carlo} algorithm},
  Physics Letters B, 268 (1991), p.~247–252.

\bibitem{Huang_2011}
{\sc W.~Huang and R.~D. Russell}, {\em Adaptive Moving Mesh Methods}, Springer
  New York, 2011.

\bibitem{Hug_2013}
{\sc S.~Hug, A.~Raue, J.~Hasenauer, J.~Bachmann, U.~Klingm{\"u}ller, J.~Timmer,
  and F.~Theis}, {\em High-dimensional {Bayesian} parameter estimation: Case
  study for a model of {J}{A}{K}2/{S}{T}{A}{T}5 signaling}, Mathematical
  Biosciences, 246 (2013), pp.~293--304.

\bibitem{Hwang_2025}
{\sc Y.~Hwang, H.~J. Kim, W.~Chang, C.~Hong, and S.~N. MacEachern}, {\em
  {Bayesian} model calibration and sensitivity analysis for oscillating
  biological experiments}, Technometrics,  (2025), pp.~1--11.

\bibitem{Kazarnikov_2020}
{\sc A.~Kazarnikov and H.~Haario}, {\em Statistical approach for parameter
  identification by {Turing} patterns}, Journal of Theoretical Biology, 501
  (2020), p.~110319.

\bibitem{Krauskopf_2007}
{\sc B.~Krauskopf, H.~M. Osinga, and J.~Gal{\'{a}}n-Vioque}, {\em Numerical
  continuation methods for dynamical systems}, Springer Netherlands, Dordrecht,
  The Netherlands, 2007.

\bibitem{Lee_2005}
{\sc S.-H. Lee, K.~Palmo, and S.~Krimm}, {\em {W}{I}{G}{G}{L}{E}: A new
  constrained molecular dynamics algorithm in {Cartesian} coordinates}, Journal
  of Computational Physics, 210 (2005), p.~171–182.

\bibitem{Leimkuhler_2015}
{\sc B.~Leimkuhler and C.~Matthews}, {\em Molecular Dynamics: With
  Deterministic and Stochastic Numerical Methods}, Springer New York, New York,
  United States, 2015.

\bibitem{Leimkuhler_2016}
{\sc B.~Leimkuhler and C.~Matthews}, {\em Efficient molecular dynamics using
  geodesic integration and solvent{\textendash}solute splitting}, Proceedings
  of the Royal Society A: Mathematical, Physical and Engineering Sciences, 472
  (2016), p.~20160138.

\bibitem{Leimkuhler_2017}
{\sc B.~Leimkuhler, C.~Matthews, and J.~Weare}, {\em Ensemble preconditioning
  for {Markov} chain {Monte} {Carlo} simulation}, Statistics and Computing, 28
  (2017), p.~277–290.

\bibitem{Lelievre_2019}
{\sc T.~Leli{\`{e}}vre, M.~Rousset, and G.~Stoltz}, {\em Hybrid {Monte} {Carlo}
  methods for sampling probability measures on submanifolds}, Numerische
  Mathematik, 143 (2019), pp.~379--421.

\bibitem{Lelievre_2022}
{\sc T.~Lelièvre, G.~Stoltz, and W.~Zhang}, {\em Multiple projection {Markov}
  chain {Monte} {Carlo} algorithms on submanifolds}, IMA Journal of Numerical
  Analysis, 43 (2022), p.~737–788.

\bibitem{Mackay_2003}
{\sc D.~J. MacKay}, {\em Information Theory, Inference and Learning
  Algorithms}, Cambridge University Press, Cambridge, United Kingdom, 2003.

\bibitem{Mangoubi_2018}
{\sc O.~Mangoubi and N.~K. Vishnoi}, {\em Dimensionally tight bounds for
  second-order {Hamiltonian} {Monte} {Carlo}}, in Proceedings of the 32nd
  International Conference on Neural Information Processing Systems, NIPS'18,
  Red Hook, NY, USA, 2018, Curran Associates Inc., p.~6030–6040.

\bibitem{Matthews_2018}
{\sc C.~Matthews, J.~Weare, A.~Kravtsov, and E.~Jennings}, {\em Umbrella
  sampling: A powerful method to sample tails of distributions}, Monthly
  Notices of the Royal Astronomical Society, 480 (2018), pp.~4069--4079.

\bibitem{Mello_2018}
{\sc B.~A. Mello, W.~Pan, G.~L. Hazelbauer, and Y.~Tu}, {\em A dual regulation
  mechanism of histidine kinase {Che}{A} identified by combining
  network-dynamics modeling and system-level input-output data}, PLOS
  Computational Biology, 14 (2018), p.~e1006305.

\bibitem{Metropolis1953}
{\sc N.~Metropolis, A.~W. Rosenbluth, M.~N. Rosenbluth, A.~H. Teller, and
  E.~Teller}, {\em Equation of state calculations by fast computing machines},
  Journal of Chemical Physics, 21 (1953), p.~1087–1092.

\bibitem{Murakami_2014}
{\sc Y.~Murakami}, {\em {Bayesian} parameter inference and model selection by
  population annealing in systems biology}, PLOS ONE, 9 (2014), p.~e104057.

\bibitem{Neal_2001}
{\sc R.~M. Neal}, {\em Annealed importance sampling}, Statistics and Computing,
  11 (2001), pp.~125--139.

\bibitem{Nesterov_2012}
{\sc Y.~Nesterov}, {\em Gradient methods for minimizing composite functions},
  Mathematical Programming, 140 (2012), p.~125–161.

\bibitem{Ottobre_2016}
{\sc M.~Ottobre, N.~S. Pillai, F.~J. Pinski, and A.~M. Stuart}, {\em A function
  space {H}{M}{C} algorithm with second order {Langevin} diffusion limit},
  Bernoulli, 22 (2016), pp.~60--106.

\bibitem{Weirs_2005}
{\sc T.~Plewa, T.~Linde, and V.~G. Weirs}, {\em Adaptive Mesh Refinement -
  Theory and Applications: Proceedings of the Chicago Workshop on Adaptive Mesh
  Refinement Methods, Sept. 3–5, 2003}, Springer Berlin Heidelberg, Berlin,
  Germany, 2005.

\bibitem{Pullen_2014}
{\sc N.~Pullen and R.~J. Morris}, {\em {Bayesian} model comparison and
  parameter inference in systems biology using nested sampling}, PLOS ONE, 9
  (2014), p.~e88419.

\bibitem{rioudurand2022metropolis}
{\sc L.~Riou-Durand and J.~Vogrinc}, {\em Metropolis adjusted {Langevin}
  trajectories: a robust alternative to {Hamiltonian} {Monte} {Carlo}}, arXiv
  preprint arXiv:2202.13230,  (2022).

\bibitem{Ryckaert_1977}
{\sc J.-P. Ryckaert, G.~Ciccotti, and H.~J. Berendsen}, {\em Numerical
  integration of the {Cartesian} equations of motion of a system with
  constraints: molecular dynamics of $n$-alkanes}, Journal of Computational
  Physics, 23 (1977), p.~327–341.

\bibitem{Smith_1987}
{\sc H.~Smith}, {\em Oscillations and multiple steady states in a cyclic gene
  model with repression}, Journal of Mathematical Biology, 25 (1987),
  p.~169–190.

\bibitem{Sokal_1997}
{\sc A.~Sokal}, {\em {Monte} {Carlo} Methods in Statistical Mechanics:
  Foundations and New Algorithms}, Springer US, 1997, p.~131–192.

\bibitem{stan2024}
{\sc {Stan Development Team}}, {\em Stan modeling language users guide and
  reference manual, version 2.36.0}, 2024, \url{http://mc-stan.org/}.

\bibitem{Swendsen_1986}
{\sc R.~H. Swendsen and J.-S. Wang}, {\em Replica {Monte} {Carlo} simulation of
  spin-glasses}, Physical Review Letters, 57 (1986), p.~2607–2609.

\bibitem{Tsitouras_2011}
{\sc C.~Tsitouras}, {\em {Runge}--{Kutta} pairs of order 5(4) satisfying only
  the first column simplifying assumption}, Computers \& Mathematics with
  Applications, 62 (2011), pp.~770--775.

\bibitem{Vyshemirsky_2008}
{\sc V.~Vyshemirsky and M.~A. Girolami}, {\em {Bayesian} ranking of biochemical
  system models}, Bioinformatics, 24 (2008), pp.~833--839.

\bibitem{Wang_2019}
{\sc S.~Wang, K.~Fan, N.~Luo, Y.~Cao, F.~Wu, C.~Zhang, K.~A. Heller, and
  L.~You}, {\em Massive computational acceleration by using neural networks to
  emulate mechanism-based biological models}, Nature Communications, 10 (2019),
  p.~4354.

\bibitem{Xu_2024}
{\sc K.~Xu and M.~Holmes-Cerfon}, {\em {Monte} {Carlo} on manifolds in high
  dimensions}, Journal of Computational Physics, 506 (2024), p.~112939.

\bibitem{Xu_2010}
{\sc T.-R. Xu, V.~Vyshemirsky, A.~Gormand, A.~von Kriegsheim, M.~Girolami,
  G.~S. Baillie, D.~Ketley, A.~J. Dunlop, G.~Milligan, M.~D. Houslay, and
  W.~Kolch}, {\em Inferring signaling pathway topologies from multiple
  perturbation measurements of specific biochemical species}, Science
  Signaling, 3 (2010), pp.~ra20--ra20.

\bibitem{Zhang_2011}
{\sc Y.~Zhang and C.~Sutton}, {\em Quasi-{Newton} methods for {Markov} chain
  {Monte} {Carlo}}, in Advances in Neural Information Processing Systems,
  J.~Shawe-Taylor, R.~Zemel, P.~Bartlett, F.~Pereira, and K.~Weinberger, eds.,
  vol.~24, Curran Associates, Inc., 2011.

\end{thebibliography}

\end{document}